\documentclass[%
 reprint,
 amsmath,amssymb,
 aps,
floatfix, ]{revtex4-2}
\usepackage{epsfig}
\usepackage{varwidth}
\usepackage{graphicx}
\usepackage{dcolumn}
\usepackage{bm}
\usepackage{braket}
\usepackage{amsmath}
\usepackage{multirow}

\begin{document}
\title{Electron capture cross sections and nuclear partition functions for $fp$-shell nuclei}
\author{Jameel-Un Nabi}
 \email{jameel@giki.edu.pk}

\author{Muhammad Riaz}%
\affiliation{%
GIK Institute of Engineering Sciences and Technology, Topi 23640,\\
Khyber Pakhtunkhwa, Pakistan.}

\date{\today}

\begin{abstract}
We present calculation of electron capture
cross sections (ECC), in the limit of zero momentum transfer, using
the pn-QRPA model in stellar matter. Towards this aim we make use of
our recently introduced recipe for estimation  of nuclear partition functions. For low momentum transfer ($q
\rightarrow 0$), the nuclear matrix elements of the $\sum \sigma
\tau^{+}$ operator provide the leading contribution to the total
cross section which we estimate using the pn-QRPA model in a
multi-shell single-particle space with a schematic interaction. Key
$fp$-shell nuclei (odd A, even-even and odd-odd) bearing astrophysical importance were selected for
the calculation of ECC in stellar environment. These $fp$-shell
nuclei play crucial rule in pre-supernova evolution of massive stars and core collapse. We further present microscopic calculation of ground and excited states Gamow-Teller strength distributions and stellar
electron capture rates on these suite of nuclei. We used two different sets of empirically determined pairing gaps to calculate the ECC and electron capture rates. Results are compared with experimental data and previous computations. Our calculated ECC are
systematically smaller at low electron incident energies as compared
to the shell-model results.
\begin{description}
\item[PACS numbers]21.10.Ma; 21.60.Jz; 23.40.-s; 26.30.Jk; 26.50.+x

\end{description}
\end{abstract}

\pacs{21.10.Ma; 21.60.Jz; 23.40.-s; 25.60.Lg; 26.30.Jk; 26.50.+x   }
\maketitle


\section{\label{level1}INTRODUCTION}
The explosive scenario in which massive stars end their lives is a well known
phenomenon called supernova. Evolution of massive stars and the
associated supernova explosion are intricately connected to the
synthesis of elements. It has been stressed with the passage of time
that calculations of stellar nucleosynthesis, neutrino-induced
reactions and nuclear aspects of supernova collapse $\&$ simulation
be based on $\textit{microscopic}$ global predictions for the nuclear
ingredients rather than on simplified semiempirical approaches.
Determination of mass of the stellar core, commencement of
gravitational collapse of the core of a massive star, thereby
triggering a supernova explosion, and fate of the resulting shock
wave, are few astrophysical events where electron capture rates play
a crucial role.

The electron capture cross section on iron group
nuclei as well as capture of electrons on free
protons play a significant role during the pre-collapse phases. It
is a well established fact that, for low stellar temperatures
(3.5$-9.3$ $GK$) and  densities ($\sim$~10$^{10}$ $gcm$$^{-3}$), electron
capture rates are sensitive to the associated Gamow-Teller (GT)
strength distribution (as the electron chemical potential is at par
with the nuclear $Q$ value). Under such conditions
electron capture mainly occurs on heavy nuclei with A $\sim$ 60. As
the stellar core stiffens to higher densities and temperatures sore
further, the electron chemical potential  exceeds the $Q$
value, and in this phase, electron capture rates are largely
dictated by the total GT strength and its centroid energy. Under
these conditions electron capture occurs on heavier nuclei A
$\ge$ 60. For even larger values of electron chemical potential
($\ge$ 20 $MeV$) corresponding to stellar densities $\ge$
10$^{11}$ $gcm$$^{-3}$, forbidden transitions become significant and should no longer be neglected. For further details and
references see \cite{Gia15}. It is therefore desirable to have a
microscopic calculation of GT strength distributions, preferably for dozens of nuclei, for a reliable prediction of electron capture
rates on $sd$-, $fp$- and $fpg$-shell nuclei. Theoretical
estimations for GT distributions fall generally into three genre:
simple independent-particle models (e.g. Ref.~\cite{Fuller});
full-scale interacting shell-model; and, in between,
the random-phase approximation (RPA) and quasi-particle random-phase
approximation (QRPA). Independent-particle models usually underestimate
the total GT strength. One reason for this is the insufficient fragmentation of the Fermi surface. At the same time these models tend to place the centroid of the GT strength too
high for  even-even parent nuclei and too low  on odd-A and
odd-odd parents \cite{Lan00}. On the other end,
full interacting shell-model
calculations are computationally taxing. In these models one may use
the Lanczos algorithm to  generate the GT strength distribution. However,
for medium-mass nuclei, one still needs to choose from among a number
of competing semi-realistic/semi-empirical interactions. RPA/QRPA models may be taken as decent approximations to a full shell-model
calculation and are  computationally much less taxing.

There is yet another challenge for the weak-interaction rate
calculations and associated nuclear models. Stellar rates take place
at temperatures of the order of a few hundred Kelvin's to a few $GK$'s. At such high temperatures the GT transitions are possible not only from the usual ground state, but
also from parent excited states. This is precisely where most of the
nuclear models compromise to approximations, e.g. Brink-Axel hypothesis \cite{Axe62} and
back resonances. Brink-Axel hypothesis states that GT strength
distribution on excited states is \textit{identical} to that from
ground state, shifted \textit{only} by the excitation energy of the
state. GT back resonances are the states reached by the GT
transitions in the electron capture direction built on
ground and excited states. The independent-particle shell model
calculation (e.g. \cite{Fuller}), spectral distribution theory (e.g.
\cite{Kar94}) and shell-model diagonalization approach (e.g.
\cite{Lan00}) are few of the models that used such approximations in
calculation of weak-rates. It is to be noted that shell model diagonalization approach estimates stellar weak rates, in a microscopic fashion, only up to first few $MeV$ of parent excitation energy. For higher parent excitation  energies  they revert back to the Brink-Axel hypothesis. Later Nabi and
Klapdor-Kleingrothaus \cite{Nab99,Nab99a,Nab04}  showed that
Brink-Axel hypothesis and back resonances are not a good approximation
for use in stellar rates. Dzhioev and collaborators \cite{Dzh10} introduced the thermofield dynamics (TFD) formalism for calculation of stellar rates that took into consideration the weak-interaction response at finite temperatures. The authors demonstrated that thermal effects shifted the GT centroid to lower excitation energy. The TFD formalism also avoided the use of Brink-Axel hypothesis. On a microscopic level handling of GT
strength distributions and associated weak-interaction rates under
stellar conditions can be tackled using two different schemes. One
(and the preferred) method is a state-by-state microscopic
evaluation of the rate summing over Boltzmann-weighted, individually
determined GT strengths for the various states (e.g. stellar weak
rates calculation using the pn-QRPA model
\cite{Nab99,Nab99a,Nab04}). The second is based on an equilibrium
statistical formulation of many-body problem (normally referred to
as shell-model Monte Carlo (SMMC) method) (e.g. \cite{Dea98}). The
SMMC  calculation found that the GT centroid shifted to lower
excitation energies and resulted in bigger widths (primarily due to
appearance of low-lying states) as temperature increases. Both the
pn-QRPA and SMMC calculations did not support the Brink-Axel
hypothesis. Despite some success, path-integral methods
are limited to interactions not plagued with the "sign problem". They are
computationally very demanding (i.e., require
supercomputer time). At low temperatures the SMMC method has few restrictions in its
applicability to odd-A and odd-odd nuclei. The pn-QRPA model can be used for any arbitrarily heavy
system of nucleons.
The thermodynamic conditions spanned over several $100 ms$ time
interval post bounce in core-collapse supernova simulation have been
analyzed \cite{Fis14}, where the density range spanned about $10$
orders of magnitude (10$^{5}$ - 10$^{15}$ $gcm$$^{-3}$), the
temperature around two orders (1 - 589 $GK$) and the lepton fraction
more than an order of magnitude (0.01 - 0.5). The large variety of
nuclear matter properties lead to distinct outcomes in supernova
simulations. The rate of change of lepton-to-baryon fraction ($Y_e$)
depends both on mass fractions and weak-interaction rates of the
nuclei.

Accurate and reliable estimates  of
nuclear partition functions (NPFs) are a prerequisite for the
determination of the electron capture cross sections. NPFs also
play key role in other astrophysical processes, e.g. determining equation of state during gravitational collapse \cite{Fuller},
mass fractions determination in the presupernova matter
\cite{Auf94} and abundance calculations during silicon burning phases of stars
\cite{Hix96}. Authors in Ref.~ \cite{Rau00}  performed computation of NPFs and reaction rates  for nuclei with
10$\le Z \le $83 using the Hauser-Feshbach
formalism. The chosen temperature range was
0.01$\le~T$~$\le$10$~GK$. Two different mass
models, namely the finite range
droplet model (FRDM) \cite{Moe95} and an extended Thomas-Fermi
approach with Strutinski integral (ETFSI) \cite{Pea96}, were employed for their calculation of NPFs. Later Rauscher \cite{Rau03} extended his estimation of NPF to extreme conditions (12 $GK \le$
T $\le $ 275 $GK$ with 9$\le Z \le $85) applying
high-temperature corrections. The authors followed a statistical approach throughout the nuclear chart. They did not  incorporate
experimental level density parameters and noted that this may lead to slightly larger deviations
from experiment. Low-lying nuclear levels need to be treated as discrete and calculated NPFs can change substantially by performing a summation over these discrete levels \cite{Dim02}.  The calculated NPF poses one of the biggest
uncertainties in the estimation of mass fractions for nuclei in presupernova cores. The calculated
NPFs displayed a variation of up to 50$\%$ when
discrete energy levels were summed as against those assuming a
level density function and performing integration over continuous states \cite{Dim02}. Inspired by this recipe, Nabi and collaborators
\cite{Nab16,Nab16a} recently calculated discrete energy levels in
700+ nuclei, up to excitation energy of 10 $MeV$, supplemented by
experimental data whenever available, and assumed a simple level
density function (for performing integration beyond 10 $MeV$) to come
up with a reliable estimation of NPFs. The same recipe for estimation of NPFs was used in
this work. We calculate  GT strength distributions using the pn-QRPA
model and NPFs in this paper. These functions were later used to estimate the electron
capture cross section on odd-A ($^{45}$Sc, $^{51}$V, $^{55}$Fe,
$^{59}$Co), even-even ($^{54,58}$Fe) and odd-odd  ($^{56,58}$Mn)
nuclei. All selected nuclei bear astrophysical significance (belonging to the
iron-regime group and  advocated to play significant role in the
presupernova evolution of massive stars). We also present
computation of associated electron capture rates for few nuclei in
this paper.

In the next section we introduce the necessary formalism
adopted in our calculation. Section III presents our estimation of
GT distribution functions, NPFs, ECC and electron capture rates.
We further compare our computation with previous calculations in
this section. Section IV finally states our conclusions.
\section{\label{level2}Theocratical frame work for calculated nuclear properties}

\subsection{The pn-QRPA formalism }

The single-particle states were calculated using a deformed Nilsson
potential \cite{Nil55}. The  proton-proton and neutron-neutron pairing
correlations were explicitly taken into account in the Bardeen-Cooper-Schriffer (BCS)
theory. We employed the BCS calculation for neutron and
proton systems independently. Ground state correlations, two-particle
and two-hole mixing states were included in the pn-QRPA model.
The ground states of parent nucleus $\ket{i}$ were computed using
the BCS equations. The
excited states of daughter nuclei $\ket{f}$ and connecting GT transitions were
calculated using the pn-QRPA equations \cite{Nab07}. The Hamiltonian was taken as
\begin{equation}
H^{pn-QRPA} =H^{sp} +V^{pair} +V_{GT}^{ph} +V_{GT}^{pp},
\end{equation}
where $H^{sp}$ is the single-particle Hamiltonian, $V^{pair}$  is the
pairing force,
$V_{GT}^{ph}$ is the particle-hole (ph) GT force and $V_{GT}^{pp}$
is the particle-particle (pp) GT force. The proton-neutron residual
interactions occurred as particle-hole and particle-particle
interaction in our model. Nilsson potential parameters were adopted from Ref.~\cite{Rag84}, Q-values were taken
from the recent mass compilation of Audi et al. \cite{Aud17} and $\hbar \omega = 41/A^{1/3}$ in units of $MeV$ was considered for
Nilsson oscillator constant (similar for neutrons and protons).  The BCS calculation
provided occupation amplitudes  and quasiparticle energies. It is to be noted that the pairing gaps ($\Delta$) were not calculated  using the BCS theory and were determined empirically in our calculation. The pairing gaps (in units of $MeV$)
were calculated using two different schemes. In the first method we chose the pairing gaps according to the global systematics \cite{Bohr}
\begin{equation}
    \Delta _{n} =\Delta _{p} =12/\sqrt{A} ,
\end{equation}
We called these pairing gap values as Scheme-I in this paper.

The paring gaps can also be determined from the empirical masses of a sequence of isotopes or isotones. This 3-point formula   is given by \cite{Bohr}
\begin{equation}
\begin{split}
\Delta n=\dfrac{1}{4}(-1)^{A-Z+1}[S_n(A-1, Z)\\-2S_n(A,Z)+S_n(A+1, Z)]
\end{split}
\end{equation}
\begin{equation}
\begin{split}
\Delta p=\dfrac{1}{4}(-1)^{Z+1}[S_p(A+1, Z+1)\\-2S_p(A,Z)+S_p(A-1, Z-1)]
\end{split}
\end{equation}
The 3-point formula was referred to as  Scheme-II in this paper. Table~1 shows the calculated values of pairing gaps using the two schemes. We investigated the effect of pairing gaps on the calculated electron capture rates and cross sections which we discuss in the next section.

Mean-field calculations can determine the deformations of nuclei (e.g., by calculating potential energy surfaces). We on the other hand determined the nuclear quadruple deformation parameter ($\beta_{2}$) using the formula
\begin{equation}
    \beta_{2} = \frac {125 (Q_{2})} {1.44 (A)^{2/3} (Z)},
\end{equation}
where $Q_{2}$ denotes the electric quadrupole moment taken from Ref.~
\cite{Mol16}.  The calculation of $\Delta$ and $\beta_{2}$ was not performed in a self-consistent way in our model. For further details of the pn-QRPA formalism and  selection of model parameters we refer to \cite{Hir93}.

\subsection{The GT strength distributions }

In the pn-QRPA model the charge-changing transitions are described by phonon
creation operators defined by
\begin{equation}
    A_{\omega}^{+}(\mu)=\sum_{pn}(X^{pn}_{\omega}(\mu)a_{p}^{+}a_{\bar{n}}^{+}-Y_{\omega}^{pn}(\mu)a_{n}
    a_{\bar{p}}).
\end{equation}
The summation was taken over all the $p-n$ pairs having $\mu$ =
(\textit{m$_{p}$ - m$_{n}$}) = 0, $\pm$1, where
\textit{m$_{n}$}(\textit{m$_{p}$}) denotes the third component of angular momentum for neutron (proton). The
\textit{a$^{+}_{n(p)}$} are the creation operator of a
quasi-particle state of neutron (proton). The \textit{$\bar{p}$}
(\textit{$\bar{n}$}) represents the time reversed state of
\textit{p} (\textit{n}).  The excitation energy ($\omega$)
and corresponding amplitudes (\textit{X$_{\omega}, Y_{\omega}$}) of phonon
operator were obtained by solving the  usual RPA equation. The ground level of our model with respect
to the QRPA phonon was taken as the vacuum,
A$_{\omega}(\mu)|QRPA\rangle$ = 0. Detailed
solution of RPA matrix equation may be seen from Refs.~\cite{Hir91, Mut89}.

Construction of parent and daughter excited states in our model may be seen from appendix A.

\subsection{Nuclear partition functions}
A novel recipe for the calculation of nuclear
partition functions (NPFs)  was
recently introduced \cite{Nab16, Nab16a}. The two key features of
this recipe were (i) treatment of low-lying nuclear states (up to 10
$MeV$) as discrete ones, and (ii) incorporation of experimental data
(energy levels and $J^{\pi}$ values) wherever possible. The authors
in Ref.~\cite{Dim02} concluded that treatment of low-lying states as
discrete can lead to substantial changes in the calculated NPF. These were computed as
\begin{equation}\label{1NPF}
\begin{split}
G(A,Z,T)=\sum_{i = 0}^{i_{m}}(2
J_{i}+1)\exp[{-E_{i}/kT}]\\
+\int_{E^{i_{m}}}^{E^{max}}\sum_{J_{i}, \pi_{i}}(2
J_{i}+1)\exp(-\epsilon/kT)\rho_{nucl}(\epsilon, J_{i},\pi_{i})d\epsilon,
\end{split}
\end{equation}
with $i_{m}$ being the label of last included experimentally known
and/or theoretically calculated parent energy state.  In
Eq.~(\ref{1NPF}), $\rho_{nucl}$ is the nuclear level density and other
symbols have their usual meanings. Above the last discrete state  an
integration was performed over the nuclear level density. We used
the Bardeen-Cooper-Schriffer (BCS) equations to construct the discrete energy levels, up to 10 $MeV$,
for all nuclide under consideration. For higher excited states we assumed a uniform Fermi gas and estimated the level density (using saddle point approximation yields and some additional
simplifications) as \cite{Boh69}
\begin{equation}\label{2NPF}
\rho_{nucl}(\epsilon, J_{i},\pi_{i}) \approx \rho_{nucl}(A,E) = \frac{\exp(2\sqrt{a E})}{4\sqrt{3}E},
\end{equation}
where $a = \frac{\pi^{2}g}{6}$ and the density of single particle
states for a nucleus having $A$ fermions is $g =
\frac{3A}{2E_F}$ where $E_F$ is the Fermi energy.
 For further details on estimation of NPFs we refer to
\cite{Nab16, Nab16a}.

\subsection{Electron capture cross sections}
The  $\beta$-decay and electron capture are the competitive
processes in opposite directions, controlling the lepton-to-baryon
fraction of stellar matter  during presupernova phases of massive
stars and given, respectively, by
\begin{equation}
(Z, N) \longrightarrow (Z+1, N-1) + e^- + \bar{\nu}_{e},
\end{equation}
\begin{equation}\label{1ECC}
e^- + (Z, N) \longrightarrow \nu_{e} + (Z-1, N+1).
\end{equation}

The nucleus $(Z,N)$ captures electron of incident energy ($w$) and
interacts weakly as shown  in Eq.~(\ref{1ECC}). The electron energy is distributed into
two parts: a part of it is absorbed by daughter nucleus to change  from initial $E_{i}$ to final state $E_{f}$ and
the remaining energy is carried out by emitted neutrino. The nuclear reaction cross section
calculation for the electron capture is governed by the
weak-interaction Hamiltonian
\begin{equation}
\widehat{H}_{\omega}=\dfrac{G}{\sqrt2}j_\mu^{lept}\widehat{J}^{\mu},
\end{equation}
where G=$G_Fcos\theta_c$, $G_F$ is the Fermi coupling constant and
$\theta_c$ is Cabibbo angle. The $j_\mu^{lept}$ and
$\widehat{J}^{\mu}$ are the leptonic and hadronic currents,
respectively, given as
\begin{equation}
j_{\mu}^{lep}=\bar{\psi}_{e}(x)\gamma_{\mu}(1-\gamma_5)\psi_{\nu}(x)
\end{equation}
\begin{equation}
\widehat{J}^{\mu}=\bar{\psi}_{p}(x)\gamma_{\mu}(1-c_{A}\gamma_5)\psi_{n}(x),
\end{equation}
where $\psi$ are the
spinor operators. In leptonic current the axial vector coupling and vector
coupling are of the same strength but in hadronic current the axial
component contains a factor of $c_{A}$ that is not equal to one.

Our main task was to estimate electron capture cross sections (ECC) of nuclear
reaction based on the calculation of nuclear transition
matrix elements between the initial $\ket{i}$ and final $\ket{f}$
nuclear states
\begin{equation}
\bra{f}|\widehat{H}_{\omega}|\ket{i}=\dfrac{G}{\sqrt2}l^{\mu}\int
d^3xe^{-i\mathbf{q.x}}\bra{f}\widehat{J_{\mu}}\ket{i},
\end{equation}
where $\mathbf{q}$ shows the three-momentum transfer and
$l^{\mu}e^{-i\mathbf{q.x}}$ are the leptonic matrix element used for
calculation of transition matrix element \cite{Paa09,Wal04}. The
nuclear transition matrix elements can be computed with the help
of Donnelly-Walecka multipolar decomposition that leads to charge,
longitudinal, transverse electric and transverse magnetic multipole
operators \cite{Paa09}. We, however, used the approximation $q
\rightarrow 0$ assuming low momentum transfer. Using this limiting
case, the transitions of the $GT_+
(=\sum\tau^+\sigma$) operator  provide the leading contribution to the
total ECC \cite{Goo80}.

The total stellar ECC on the target
nucleus of mass number $A$ and charge number $Z$, as a function of
incident electron energy ($w$) and temperature ($T$), is given by
\begin{equation}\label{ECC}
\begin{split}
\sigma(w,T)=\dfrac{G_F^{2}cos^2\theta_c}{2\pi}\sum\limits_{i}F(Z,w)\dfrac{(2J_{i}+1)\exp{(-E_i/kT)}}{G(A,Z,T)}\\
\times
\sum\limits_{J,f}(w-Q+E_i-E_f)^{2}\dfrac{|\bra{i}\hat{O}_J\ket{f}|^2}{(2J_{i}+1)},
\end{split}
\end{equation}
where $F(Z,w)$ is the  Fermi function  and
accounts for the distortion of the electron's wave function due to
the Coulomb field of the nucleus and finite size of the nucleus. The Fermi function was calculated using the prescription of Ref.~\cite{Gov71}. The $G(A,Z,T)$ are the NPFs and their computation was discussed in the previous section. The $J_{i}$'s are the
angular momenta of parent states. The $\hat{O}_J$ operator,
appearing in the nuclear matrix elements of Eq.~(\ref{ECC}), reduces to the
$GT_{+}$ operator (which changes a proton into a neutron) for low
three-momentum transfer. We quenched the GT strength by a
constant factor of 0.6 \cite{Gaa83}.

\subsection{Electron capture rates}
The electron capture (EC)  rate of a transition from the $i^{th}$
state of the parent to the $f^{th}$ state of the daughter nucleus is
given by
\begin{eqnarray}\label{rates}
\lambda ^{^{EC} } _{if} =\left[\frac{\ln 2}{D}
\right]\left[f_{if} (T,\rho ,E_{F} )\right] \nonumber \\
\left[B(F)_{if} +\left({\raise0.7ex\hbox{$ g_{A}
$}\!\mathord{\left/ {\vphantom {g_{A}  g_{V} }} \right.
\kern-\nulldelimiterspace}\!\lower0.7ex\hbox{$ g_{V}  $}}
\right)^{2}_{eff} B(GT)_{if} \right].
\end{eqnarray}
The value of constant $D$ was taken to be 6143s \cite{Har09}. $B_{if}'s$ are
the sum of reduced transition probabilities of the Fermi and Gamow-Teller transitions given by
\begin{equation}
    B(F)_{if} = \frac{1}{2J_{i} +1} \langle{f}\parallel\sum\limits_{k}
    \tau^{+}_{k}\parallel {i}\rangle|^{2}
\end{equation}
\begin{equation}
    B(GT)_{if} = \frac{1}{2J_{i} +1} \langle{f}\parallel\sum\limits_{k}
    \tau^{+}_{k}\overrightarrow{\sigma}^{k}\parallel {i}\rangle|^{2}.
\end{equation}

The $f_{if}$ is the phase
space integral over total energy and for electron capture it is
given by (using natural units $\hbar = c = m_{e}
=1$):
\begin{equation}\label{PHS}
 f_{if} \, =\, \int _{w_{l} }^{\infty }w\sqrt{w^{2} -1}
 (w_{m} \, +\, w)^{2} F(Z,w)\Lambda_{-} dw.
\end{equation}
In the above equation $w$ is the incident electron energy, $w_{l}$ is the total capture threshold
energy (rest+kinetic) for electron capture, $w_{m}$ is the total decay
energy
\begin{equation}
w_{m} = m_{p}-m_{d}+E_{i}-E_{f},
\end{equation}
where $m_{p}$ and $E_{i}$ are mass and excitation energies of the
parent nucleus, and $m_{d}$ and $E_{f}$ of the daughter nucleus,
respectively. $F(Z,w)$ are the Fermi functions and $\Lambda_{-}$ is the Fermi-Dirac distribution function for
electrons
\begin{equation}\label{G1}
 \Lambda_{-} =\left[\exp \left(\frac{E-E_{F} }{kT}
 \right)+1\right]^{-1}.
\end{equation}
Here $E = (w-1)$ is the kinetic energy of the electrons, $E_{F}$ is
the Fermi energy of the electrons and $k$
is the Boltzmann constant.

The number density of electrons associated with protons and nuclei
is $\rho Y_{e}N_{A}$ ($\rho$ is the baryon density and $N_{A}$ is
Avogadros number) where,
\begin{equation}\label{roh}
\rho Y_{e}=\frac{1}{\pi ^{2} N_{A} }(\frac{m_{e}c}{\hbar})^{3}\int
_{0}^{\infty }(\Lambda_{-}  -\Lambda_{+} )p^{2} dp,
\end{equation}
here $p = (w^{2}-1)^{1/2}$ is the electron momentum and Eq.~(\ref{roh}) has
the units of $mol \hspace{0.1in} cm^{-3}$. $\Lambda_{+}$ is the
Fermi-Dirac distribution function for positrons,

\begin{equation}\label{G2}
\Lambda_{+} =\left[\exp \left(\frac{E+2+E_{F} }{kT} \right)+1\right]^{-1}.
\end{equation}
We note that Fermi-Dirac distribution functions were used in
Eqs.~(\ref{G1}) and (\ref{G2}) as the electrons were assumed to be in continuum
state in stellar matter. Eq.~(\ref{roh}) was used for an iterative
calculation of Fermi energies for selected values of $Y_{e}$ and $T$.
There is a finite probability of occupation of parent excited states
in the stellar environment as a result of the high temperature in the
interior of massive stars. EC rates then also have a finite
contribution from these excited states. The total electron capture rate per unit time per nucleus was then
computed using
\begin{equation}\label{Trates}
\lambda^{EC} =\sum _{if}P_{i} \lambda _{if}^{EC}.
\end{equation}
The summation over all initial and final states was carried out until satisfactory convergence in the EC rate was
achieved in our calculation. In Eq.~(\ref{Trates}) $P_{i}$ is the probability of occupation of parent
excited states and follows the normal Boltzmann distribution.

\section{\label{level3}RESULTS AND DISCUSSION}
\begin{figure}[h]
\includegraphics [width=8.in]{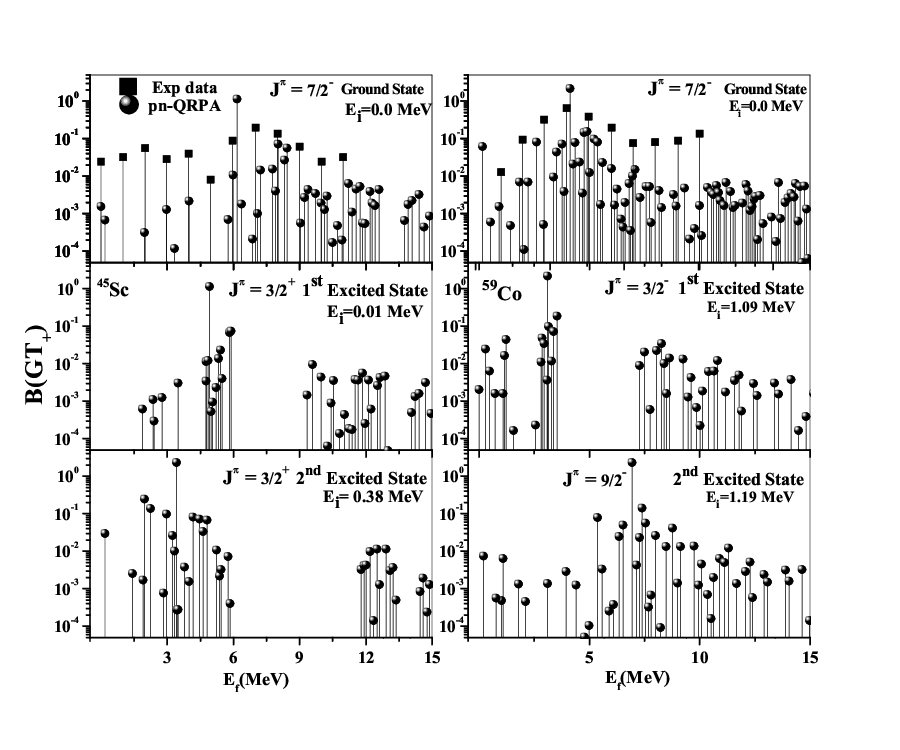}
\centering \caption{GT$_{+}$ strength distributions calculated
using the pn-QRPA theory for the ground and first two excited states
of $^{45}$Sc and $^{59}$Co. Experimental data were taken from Refs.~\cite{Fre65, Alf93}.}\label{scbgt}
\end{figure}

\begin{figure}
\includegraphics [width=8.in]{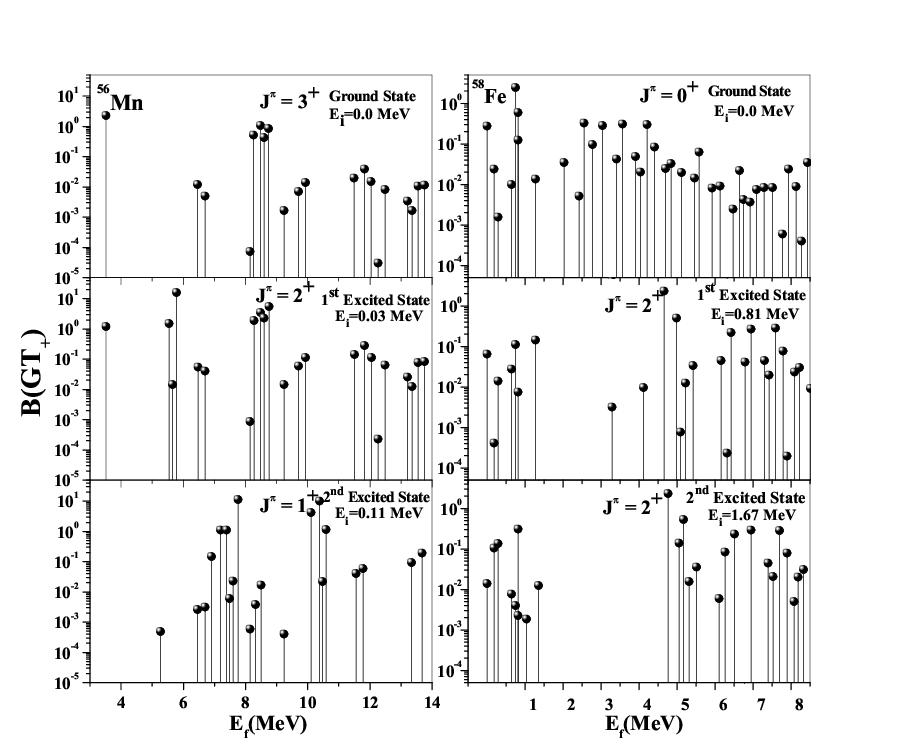}
\centering \caption{Same as Fig.~\ref{scbgt} but for
$^{56}$Mn and $^{58}$Fe.}\label{mnbgt}
\end{figure}

We start the proceedings by presenting the calculated GT strength
distributions using the pn-QRPA model. The
pn-QRPA model estimates ground and excited states GT strength
distributions in a \textit{microscopic} fashion. In other words the Brink-Axel hypothesis was not assumed in the calculation of excited states GT distributions.
Figs.~\ref{scbgt}-\ref{mnbgt} show the computed GT strength
distributions for the ground and first two excited states of
$^{45}$Sc, $^{56}$Mn, $^{58}$Fe and $^{59}$Co, respectively.  The
abscissa shows the excitation energy (in units of $MeV$) in daughter
nuclei. As mentioned earlier an overall quenching factor of 0.6
\cite{Gaa83} was adopted in the estimation of GT strength  for all
nuclei. Calculated GT transitions of magnitude less than $10^{-4}$ are
not shown in these figures. In Fig.~\ref{scbgt}, for ground states, the experimental data is shown by squares. The pn-QRPA  transitions are shown by circles. The top-left panel of Fig.~\ref{scbgt} shows a prominent peak from ground
state of parent ($^{45}$Sc) of magnitude 1.1 at 6.2 $MeV$ in
$^{45}$Ca. The peak strength shifts to a lower excitation energy of
4.9 (3.4) $MeV$ for the first (second) excited state. The right panels of Fig.~\ref{scbgt} show that the total GT strength from ground state of $^{59}$Co is much bigger than those from excited states. For the case of
$^{56}$Mn (see left panels of Fig.~\ref{mnbgt}) the low-lying GT peaks disappear
for the second excited state. Similarly right panels of Fig.~\ref{mnbgt} show that
the ground-state GT strength distribution for $^{58}$Fe  is well
populated, not to be seen for the first and second excited states. The GT distributions are also different for high-lying excited states.  Complete set of GT strength distribution
functions for higher excited states may be requested from the
authors. Fig.~\ref{scbgt} and Fig.~\ref{mnbgt} clearly show
that Brink-Axel hypothesis is not a good approximation to be used for the
calculation of stellar ECC and EC rates.

Tables~\ref{npf1}~--~\ref{npf2} show our computed NPFs as a
function of stellar temperature ($GK$) for few $fp$-shell nuclei of astrophysical importance. We further compare our results with the  statistical model calculation of Refs.~\cite{Rau00, Rau03}.  The finite range droplet model (FRDM)
\cite{Moe95} was used as one of the input mass models to calculate
the NPFs by Ref.~\cite{Rau00}. It is to be noted that the
estimations of NPFs by Refs.~\cite{Rau00, Rau03} were normalized to the
ground-state spin multiplicity whereas our NPFs include the
ground-state spin multiplicity factor. We compare our estimated NPFs with those calculated by
Ref.~\cite{Rau00} up to a temperature of 10 $GK$. At 30 $GK$ we compare our
NPFs with the computation of Ref.~\cite{Rau03} where high-temperature
corrections were incorporated by Rauscher. At low
temperatures ($\ge$ 0.5 $GK$) we do notice significant differences in the two calculations, specially for
nuclei where angular momentum of ground-state  differs significantly
from zero. We incorporated experimental energy levels (along
with their spins) in calculation of our NPFs missing in the statistical model. The NPFs
computed by Rauscher are  orders of magnitude bigger at 30 $GK$ and higher temperatures. This we attribute to incorporation of
high-temperature correction effects in Rauscher's computation which
was missing in our estimation. This we take as a drawback in our calculation.

\begin{figure}
\includegraphics [width=7.in]{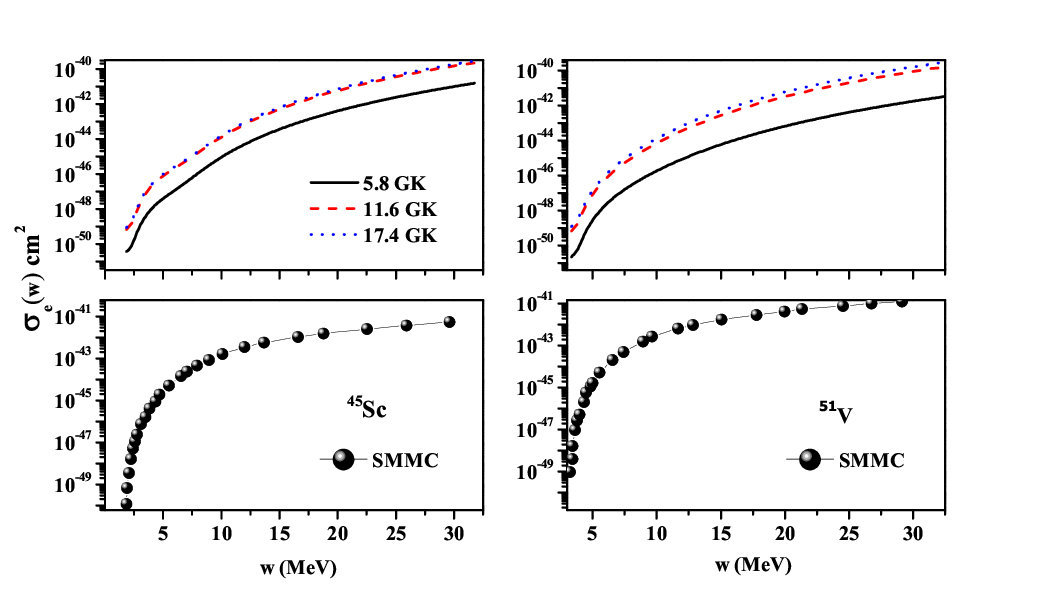}
\centering \caption{Electron capture cross sections for $^{45}$Sc and $^{51}$V,
using the pn-QRPA theory, as a function of the incident electron
energy ($w$) at different stellar temperatures. Bottom panels
show result of SMMC calculation \cite{Dea98}.}\label{sccs}
\end{figure}

\begin{figure}
\includegraphics [width=7.in]{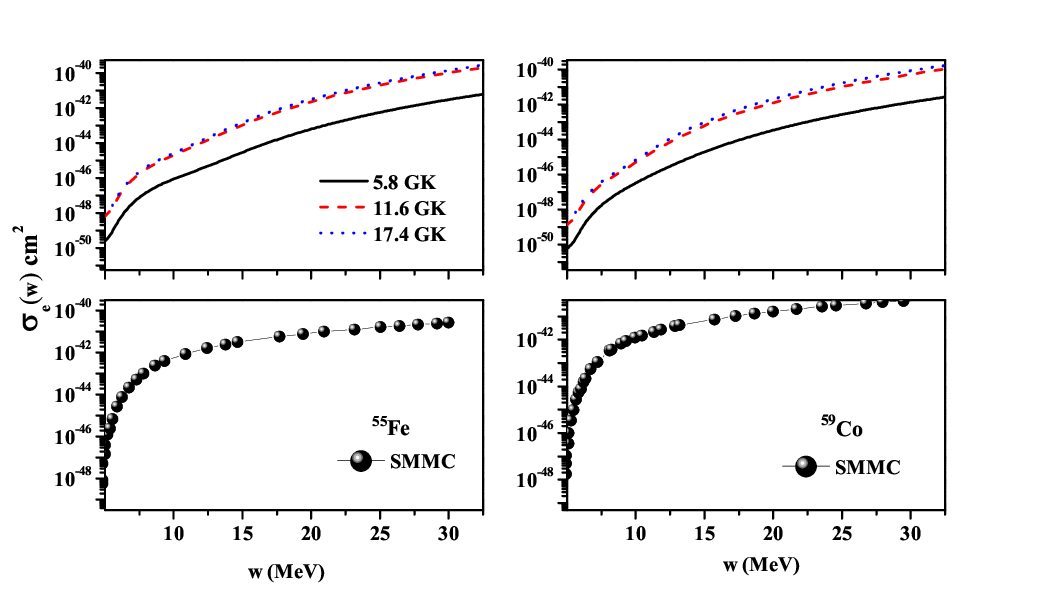}
\centering \caption{Same as Fig.~\ref{sccs} but for
$^{55}$Fe and $^{59}$Co.}\label{fe55cs}
\end{figure}

\begin{figure}
\includegraphics [width=7.in]{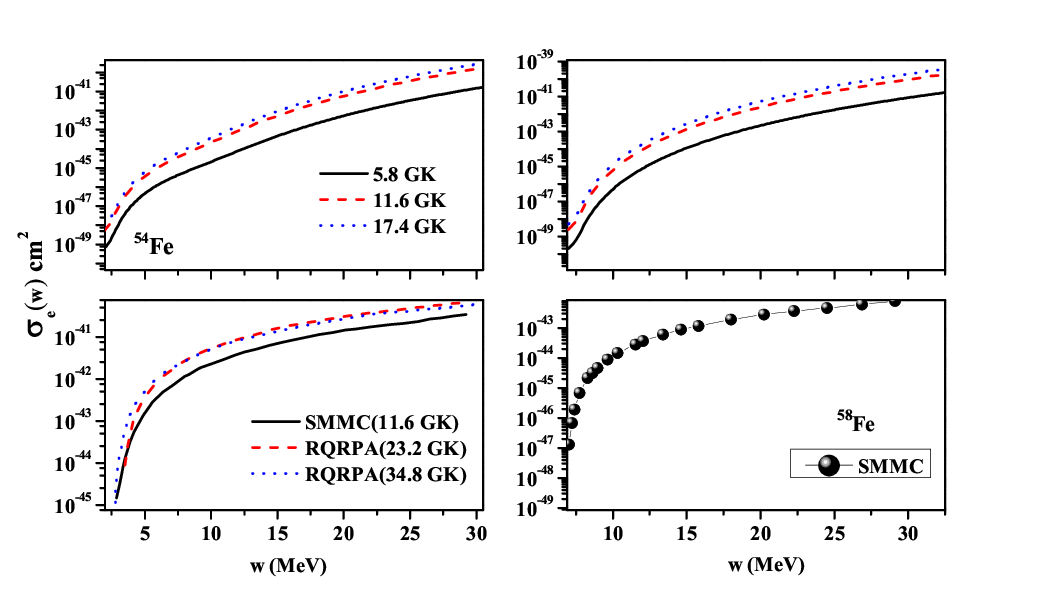}
\centering \caption{Same as Fig.~\ref{sccs} but for $^{54, 58}$Fe.
Bottom panel shows results of SMMC  \cite{Dea98} and FTRRPA
calculations  \cite{Niu11}.}\label{fe54cs}
\end{figure}

\begin{figure}
\includegraphics [width=5.in]{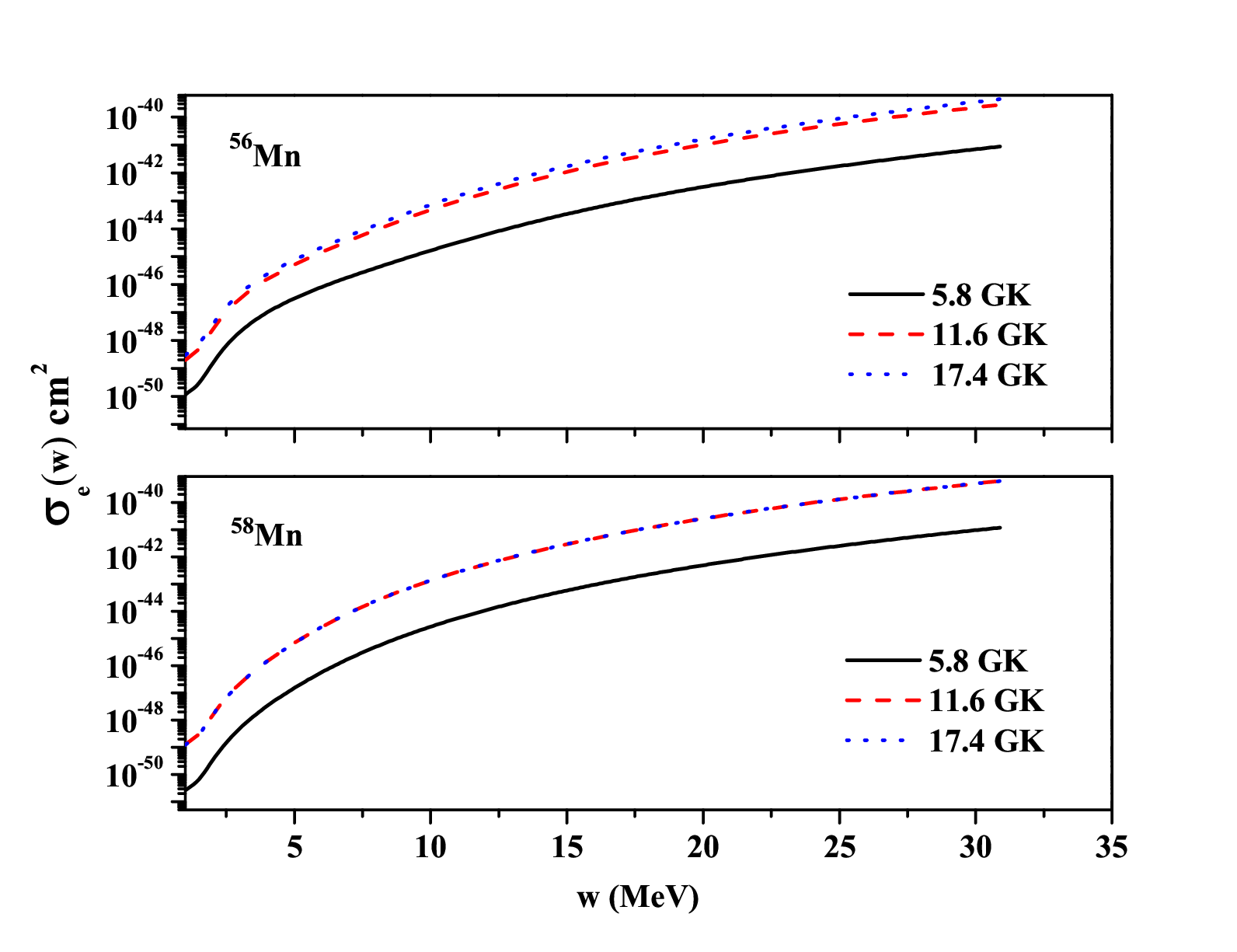}
\centering \caption{Pn-QRPA calculated electron capture cross
sections for $^{56}$Mn (top panel) and $^{58}$Mn (bottom panel) as a function of the incident electron energy
($w$) at different stellar temperatures. }\label{mncs}
\end{figure}

Matrix elements of the $\sum \sigma \tau^{+}$ operator and NPFs
are two key microscopic ingredients in the calculation of stellar
ECC. After displaying our results for the estimation of GT strength
distributions and NPFs, we now present our calculation of ECC for
key $fp$-shell nuclei in stellar environment. We
calculated our ECC for stellar temperatures (5.8 -
17.4 $ GK$). At these temperatures the GT transitions are
unblocked as the residual interaction becomes sufficiently strong. The allowed GT$_+$ transitions are Pauli blocked at low temperature due to closed neutron $f$ sub-shell. As the temperature increases unblocking of closed shells take place. Similarly GT$_+$ transition can undergo by thermally excited protons. The important range
of the incident electron energy ($w$) is up to 30 $MeV$
\cite{Gia15}. Accordingly we present our calculated ECC in
Figs.~\ref{sccs}~-~\ref{mncs} up to $w$ = 30 $MeV$.

Left panels of Fig.~\ref{sccs} shows our estimated ECC for odd-A nucleus
$^{45}$Sc as a function of incident electron energy. We calculated the ECC at
stellar temperatures of 5.8, 11.6 and 17.4 $GK$, respectively. The
minimum electron energy to initiate the electron capture process is
given by Eq.~(32). However at finite
temperatures, this threshold is further lowered by the internal
excitation energy. The ECC increases drastically within the first
few $MeV$'s of $w$ above threshold. This effect maybe traced
to the behavior of calculated GT distribution. The bottom-left panel of
Fig.~\ref{sccs} reproduces the shell model Monte Carlo computation
of ECC at a temperature of 5.8 $GK$ \cite{Dea98}. At low $w$
($\sim$ 2.5 $MeV$) our calculated ECC is around three orders of
magnitude smaller. At low incident electron energies the capture
process is sensitive to the details of GT strength distribution. The GT strength distributions of
ground and first two excited states of $^{45}$Sc were presented
earlier in Fig.~\ref{scbgt}.  The SMMC model predicts marginally
larger cross sections at low $w$. This result is expected due to
the strong configuration mixing in SMMC calculation (also see Ref.~\cite{Niu11} for further discussion). For electron energy $w
\ge$ 10 $MeV$ the estimated cross sections display a more gradual
increase. At high incident electron energy of 29.5 $MeV$ the reported
ECC is roughly factor two larger. It is to be noted that at low
$w$, the calculated ECC are very small numbers and can change
by orders of magnitude with a rather small change in the computed GT strength
distributions and NPFs. Moreover these numbers are much smaller
than those at $w \ge$ 10 $MeV$ and consequently the total electron capture
rates calculated by the two models may not be significantly
different in the low-energy interval. It may however depend strongly
on prevailing stellar density and temperature values (see related
discussion in Refs.~\cite{Paa09,Dea98}).  With increasing $w$ the ECC
continues to rise modestly (because of the $(w-Q+E_i-E_f)^{2}$
factor in Eq.~(27)). In the SMMC estimation only the $0\hbar\omega$
GT transition strength was considered, rather than the total
strength in the 1$^{+}$ channel. As the temperature increases from
5.8 $GK$ to  11.6 $GK$, we witness a notable increase up to an order of
magnitude in the calculated ECC and corresponds to an appreciable thermal unblocking of the GT$_{+}$ channel as discussed earlier. We note that
the Fermi contribution to the total ECC is negligible. Because
these transitions are already unblocked at T = 5.8 $GK$, a further
increase in temperature to T = 17.4 $GK$ results in a comparatively small increase (less than a factor two) of the calculated ECC.    Fig.~\ref{sccs} (right panel) and Fig.~\ref{fe55cs} show a similar
temperature dependence of pn-QRPA model calculated ECC and its comparison with SMMC
results  for   odd-A nuclei $^{51}$V, $^{55}$Fe and
$^{59}$Co, respectively. For all remaining cases the SMMC numbers
are roughly an order of magnitude bigger at high $w$. We note
that the SMMC model invoked a quenching factor of 0.8 in their ECC
estimation to be compared with a tighter pn-QRPA quenching
factor of 0.6.

Next we show the calculation of ECC for even-even nuclei. Several
nuclear models, including the SMMC \cite{Dea98}, the pn-QRPA model
with a Bonn C-D potential \cite{Gia15}, finite temperature HF+RPA
\cite{Paa09}, finite temperature relativistic RPA \cite{Niu11},
thermal quasiparticle RPA \cite{Dzh10} and the hybrid SMMC+RPA model
\cite{Lan01}, were used in the past, to perform ECC calculation of
even-even nuclei. For the case of $^{54}$Fe, we compare our
calculated ECC with the SMMC result \cite{Dea98} and the finite
temperature relativistic RPA (FTRRPA) estimation performed by Niu
and collaborators \cite{Niu11} in the bottom-left panel of
Fig.~\ref{fe54cs}. The FTRRPA calculation was performed up to
temperature of 23.2 $GK$. We notice that the electron capture
threshold energy remains more or less same in SMMC  (at T = 5.8 $GK$)
and FTRRPA  (at T= 23.2 $GK$) computations. The FTRRPA calculated ECC
becomes almost independent of temperature at $w \ge$ 5 $MeV$ . We
notice that the FTRRPA computed ECC is a factor 2--3 bigger than
the SMMC results. The pn-QRPA calculated threshold is somewhat
lower. At low $w$, our ECC is 2--3 orders of magnitude smaller
than FTRRPA results but at high $w$ the two estimations are in
good comparison. The SMMC results are roughly a factor three bigger
at high $w$. Right panels of Fig.~\ref{fe54cs} compare our ECC calculation with
those of SMMC for the even-even nucleus $^{58}$Fe. We note that at $w$ = 29.1 $MeV$ the pn-QRPA computed ECC is
roughly a factor 8 bigger than SMMC despite a much tighter  quenching
factor used in our calculation. The most probable reason for this
enhancement could be that SMMC computation is carried out in the
$0\hbar\omega$ $fp$-shell space. On the other hand we perform our
pn-QRPA calculation in a  $7\hbar\omega$ model space so that for any
multipole operator the whole sum rule is exhausted, whereas,
generally speaking, this is not the case in SMMC calculation.

Fig.~\ref{mncs} depicts our ECC estimation for odd-odd
nuclei $^{56,58}$Mn. The upper panel of Fig.~\ref{mncs} shows our estimated ECC for $^{56}$Mn. At
low $w$ of 2.5 $MeV$ the ECC for $^{56}$Mn is around a factor 4
bigger than for $^{58}$Mn (lower panel) for $T=$ 5.8 $GK$.
The temperature evolution is very much
similar to that for other cases. The estimated GT strength
distribution for ground and first two excited states of $^{56}$Mn
were presented earlier in Fig.~\ref{mnbgt}. The NPFs for $^{56,58}$Mn used in the calculation can be seen in Table~\ref{npf1}.

Table~IV investigates how the two different values of pairing gaps, determined in our model, effect the calculated ECC. Here we show the calculated ECC for all nine $fp$-shell nuclei using Scheme-I and Scheme-II. The calculated ECC for the two schemes are shown as a function of incident electron energy.  We note that the calculated ECC do not differ significantly for the two schemes. The Scheme-I ECC were bigger at most by a factor 2.5 for the case of $^{58}$Mn at $w$ = 2 $MeV$ while the Scheme-II ECC were at most a factor 1.5 bigger for the case of $^{56}$Mn at $w$ = 10 $MeV$. At low values of $w$ the ECC using Scheme-1 is slightly bigger  (except for the case of $^{54}$Fe). However as $w$ increases the two calculations get in better agreement. Due to a small number of nuclei investigated in this study, we are not in a position to determine which set of pairing gaps gives better results. We intend to investigate this more as a future assignment.

\begin{figure}
\includegraphics [width=5.5in]{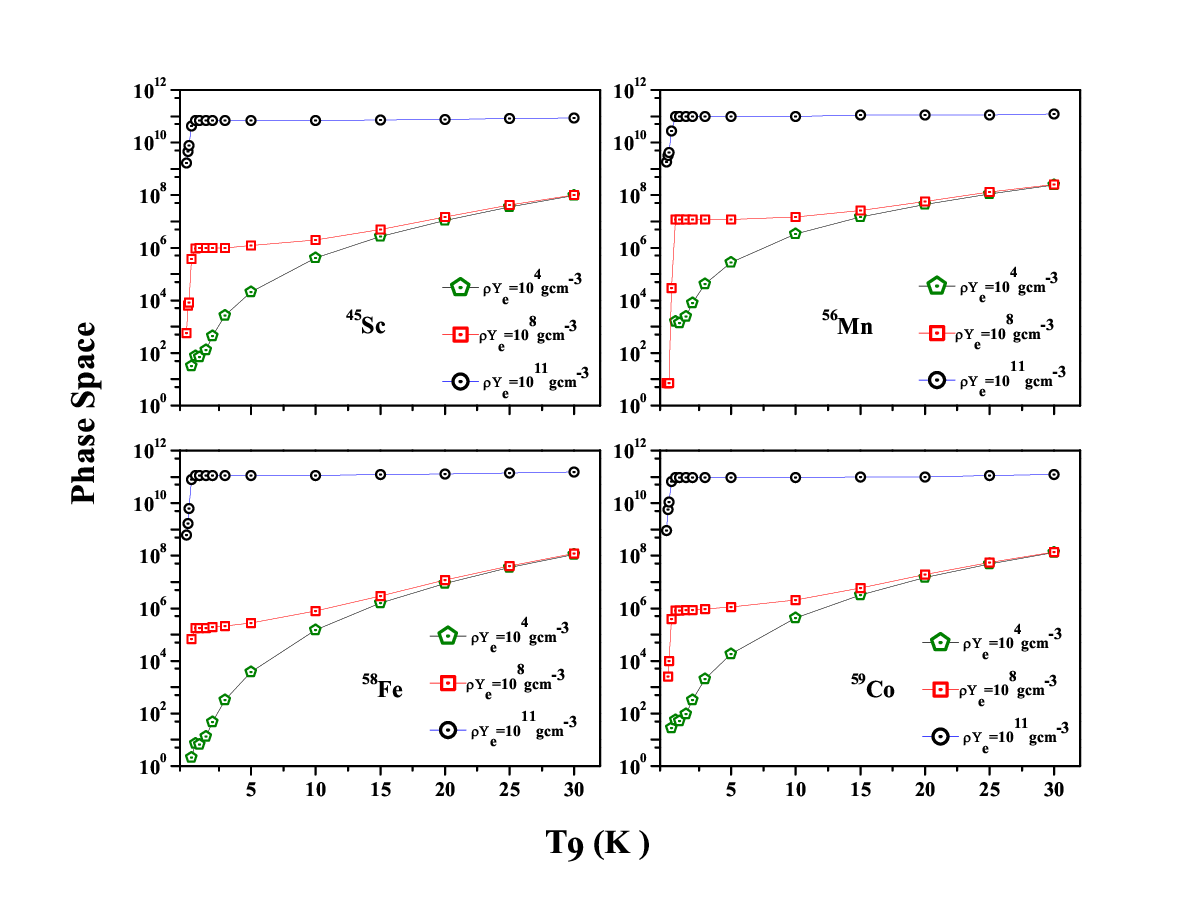}
\centering \caption{Total phase space for few $fp$-shell nuclei
as a function of stellar temperature and density.} \label{ps}
\end{figure}

We next present some results of stellar electron capture rates
for $fp$-shell nuclei. To this purpose we earmark three values of
stellar density (10$^{4}$ $g/cm$$^{3}$, 10$^{8}$ $g/cm$$^{3}$ and
10$^{11}$ $g/cm$$^{3}$) representing low, medium and regions of high
stellar density, respectively.  We adopted the prescription of
Martin and Bleichert-Toft \cite{Mar70} for the calculation of phase
space functions for electron capture. We present total phase space of $^{45}$Sc, $^{56}$Mn, $^{58}$Fe and $^{59}$Co as a
function of stellar temperature in Fig.~\ref{ps}. Each panel shows
three curves representing the three selected stellar densities. The
phase space increases with increasing stellar temperature and
density. For low and intermediate densities the phase space is
essentially zero at small stellar temperature of 0.01 $GK$ and
increases with increasing temperatures. For high density regions,
the phase space is $\sim$ 10$^{9}$ at 0.01 $GK$ and increases
to  $\sim$ 10$^{11}$ at 30 $GK$.

\begin{figure}
\includegraphics [width=6.5in]{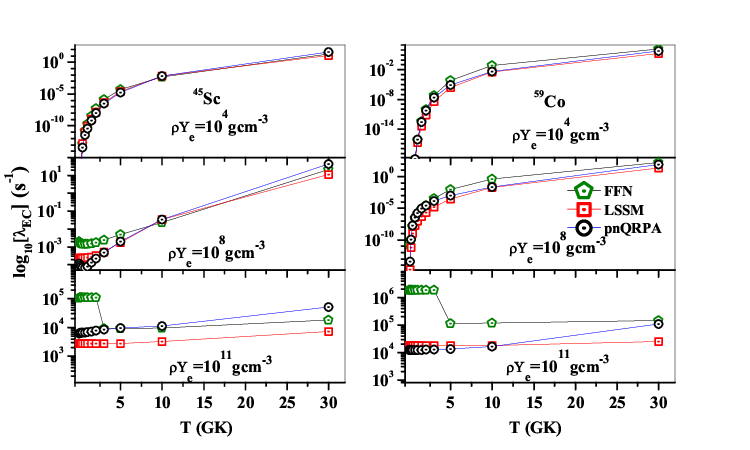}
\centering \caption{Comparison of pn-QRPA electron capture rates on
$^{45}$Sc and $^{59}$Co with those calculated by LSSM and IPM as a function of
stellar temperature and density.} \label{scec}
\end{figure}

The pn-QRPA calculated stellar electron capture rates for $^{45}$Sc and $^{59}$Co is presented in Fig.~\ref{scec}. Shown also are the results of large
scale shell model  (LSSM) \cite{Lan00} and the independent particle
model  (IPM) \cite{Fuller} calculations. The comparison is presented
over three panels depicting low, medium and high stellar densities,
respectively. At low stellar density the three computations are in
excellent agreement. At 30 $GK$, our calculated rates are 41
$s^{-1}$, to be compared with 20 $s^{-1}$ (IPM) and 10 $s^{-1}$
(LSSM). At intermediate density, pn-QRPA rates are slightly smaller
at low temperatures. At 30 $GK$, we have more or less the same
comparison. At high density and low temperature IPM results are
around two orders of magnitude bigger than pn-QRPA and LSSM rates.
At 30 $GK$, the pn-QRPA calculation is around factor 3 bigger
than IPM rates. LSSM rates are again small at high temperatures. The
Lanczos-based approach employed in LSSM calculation may explain the source of these small rates. These were first  highlighted by Pruet and Fuller~ \cite{Pru03}. The Lanczos iterations essential
for convergence in rate calculation and the corresponding behavior of partition functions may get affected
and consequently the LSSM rates tend to be too low at high temperatures. The pn-QRPA
calculation, not Lanczos-based in nature, do not suffer from this convergence problem. At high temperatures, where the occupation
probability of excited states is larger, our EC rates are
accordingly enhanced compared to LSSM and IPM rates.

\begin{figure}
\includegraphics [width=6.5in]{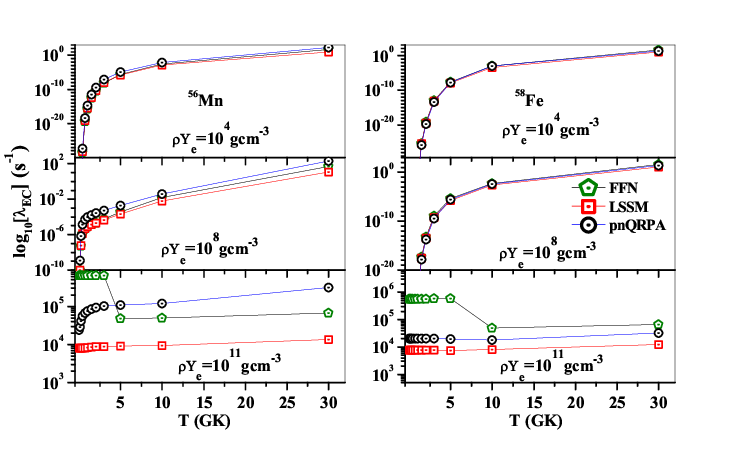}
\centering \caption{Same as Fig.~\ref{scec} but for $^{56}$Mn and $^{58}$Fe.}
\label{mnec}
\end{figure}

The left panel of Fig.~\ref{mnec} shows pn-QRPA estimated electron capture rates and
comparison with IPM and LSSM rates on
$^{56}$Mn. At low stellar density the comparison with previous
calculations is akin to the case of $^{45}$Sc. At intermediate
density the pn-QRPA rates are around one order of magnitude bigger
at all temperatures whereas at high density the comparison is once
again similar to what we witnessed for the case of $^{45}$Sc (for further detail see Ref.~\cite{Maj18}).
Unmeasured matrix elements  were assigned an
average value of $log ft = $5 in IPM calculation for allowed transitions. On the other hand
these transitions were computed in a microscopic fashion using the
pn-QRPA and LSSM approaches  thereby increasing the reliability of the  calculated rates.

We finally depict the electron capture rates for the even-even nucleus $^{58}$Fe and comparison with previous calculations in
right panel of Fig.~\ref{mnec}.  The LSSM rates are around a factor 2 -- 3
smaller at high temperatures for reason already mentioned. At high
stellar density the IPM rates are bigger for all temperatures. IPM
threshold parent excitation energies were not constrained and
extended well beyond the particle decay channel taken into account in the current model. At high
temperatures contributions from these high excitation energies begin
to show their cumulative effect. The IPM rates are usually
bigger than the pn-QRPA and LSSM rates, specially at high
stellar density. The misplacement of GT centroid (for further details on
centroid placement in LSSM and IPM calculations see Refs.~\cite{Lan00, Lan001}),  assignment of an
average value of $log ft = $5 for all unmeasured transitions and performing sum on parent excitation energies  well beyond the particle decay channel could be the probable reasons for this enhancement in IPM rates. The pn-QRPA model provides an adequate
model space that can effectively handle all the excited states in
parent as well as in daughter nuclei. We do not consider the
Brink-Axel hypothesis in our calculation to approximate the EC rate
contribution from parent excited levels. This approximation was used
both by IPM and LSSM. Accordingly, whenever  rates from ground state
command the total rate, our electron capture rates are in decent
agreement with previous calculations. For cases where excited state
partial rates influence the total rate, differences are seen between
the pn-QRPA and previous results.

The influence of pairing gap values on EC rates is explored in Table~V. Here we calculate EC at fixed stellar densities of 10 $g/cm$$^{3}$, 10$^{4}$ $g/cm$$^{3}$ and 10$^{8}$ $g/cm$$^{3}$ and core temperatures of 5 $GK$, 25 $GK$ and 30 $GK$ for the two pairing gap schemes for all nine nuclei. Once again we notice that there is no significant change in the calculated EC values for the two different schemes. As the core temperature increases we note that the EC rates calculated using pairing gap values of Scheme-II increase. We are currently working on the role of pairing correlations in calculation of $\beta$-decay half-lives  \cite{Nab19} where we investigate in detail which scheme results in a better prediction of calculated half-life values using the pn-QRPA model. Our preliminary results favor the use of Scheme-II.

\section{\label{level4}SUMMARY AND CONCLUSION}

We presented a microscopic calculation of  GT strength distributions for key
$fp$-shell nuclei from ground and excited states of the parent nucleus.  Our results were indicative of the fact that Brink-Axel hypothesis may not be a good
approximation to be used for the estimation of stellar weak
rates. The computed NPFs emphasized on the
treatment of nuclear excited states as \textit{discrete} up to 10
$MeV$. The energy levels were calculated
by solving the relevant BCS equations which gave us the quasiparticle
energies and probabilities for each particle level to be
(un)occupied. The resulting NPFs showed considerable differences when compared with those calculated using traditional recipes. \\
We calculated ECC for even-even, odd-A and odd-odd $fp$-shell nuclei. We used the pn-QRPA model in a multi-shell
single-particle space with a schematic interaction to estimate ECC
for $fp$-shell nuclei of astrophysical significance. The temperature
dependence of ECC was explored. As the temperature increased from 5.8 $GK$ to 11.6
$GK$, we calculated a noticeable increase, up to an order of magnitude,
in the ECC corresponding to a significant thermal unblocking of the
GT$^{+}$ channel. There was a marginal increase in
the computed ECC when the core temperature was raised further. At low
incident electron energies, our calculated ECC was systematically
smaller than SMMC results which we attribute to the strong
configuration mixing in SMMC calculation and a tighter quenching factor adopted in our model. For high incident energies shell model results are factor eight (two) smaller for $^{58}$Fe ($^{45}$Sc).\\

The pn-QRPA calculated electron capture
rates for $fp$-shell nuclei in stellar matter were shown. These were, in general, found to be bigger
than previous estimations at high stellar temperatures. The ASCII
files of all computed stellar rates, on a fine temperature-density
grid suitable for interpolation purposes, are available and may be
requested from the authors.

The role of empirical pairing gap for protons and neutrons on calculated ECC and EC rates was investigated. In first instance the pairing gaps were taken to be the same for protons and neutrons (Scheme-I). Later we used pairing gaps using 3-point formula (Scheme-II). The latter are argued to be more accurate.  The calculated EC rates using Scheme-II were slightly bigger at high core temperatures. For low incident electron energies the calculated ECC is slightly bigger using pairing gaps from Scheme-I. As $w$ increases the two calculations get in better agreement.

The ECC calculation and associated EC rates presented in this paper have many astrophysical
applications. They can be used for estimating neutrino-spectra
arising from electron capture on nuclei during supernova phase. One
needs the neutrino-spectra in the stellar core, as a function of
position and time, for modeling and simulation of the presupernova
and supernova phases of massive stars. The energy loss rates due to
supernova neutrinos may  determine a specific supernova-neutrino
scenario. One can estimate the neutrino emissivity using the
neutrino-spectra (see e.g. \cite{Lan00}). The inverse neutrino
absorption process can  be determined using the EC
rates presented in this work and would be taken as a future
assignment. In future we plan to investigate role of forbidden
transitions in calculation of ECC, specially at high electron
incident energies. \vspace{0.5in}

J.-U. Nabi would like to
acknowledge the support of the Higher Education Commission
Pakistan through project numbers  5557/KPK/NRPU/R$\&$D/HEC/2016 and  9-5(Ph-1-MG-7)Pak-Turk/R$\&$D/HEC/2017 and
Pakistan Science Foundation through project number
PSF-TUBITAK/KP-GIKI (02).

\section{\label{level5}Appendix A}
Excited states of even-even nucleus were obtained by invoking
one-proton (or one-neutron) excitations. They were described, in the
quasiparticle (q.p.) picture, by adding two-proton (two-neutron)
q.p.'s to the ground state \cite{Mut92}.
Allowed transitions from these initial states to final
proton-neutron q.p. pair states in the odd-odd daughter nucleus were calculated keeping in mind the selection rules. The
transition amplitudes and their reduction to correlated ($c$)
one-q.p. states \cite{Mut92} were given by

\begin{equation}\tag{1A}\label{first}
\begin{split}
<p^{f}n_{c}^f \mid \tau^{+}\sigma_{-\mu} \mid p_{1}^{i}p_{2c}^{i}>
 = -\delta (p^{f},p_{2}^{i}) <n_{c}^{f} \mid \tau^{+}\sigma_{-\mu} \mid p_{1c}^{i}>\\
+\delta (p^{f},p_{1}^{i}) <n_{c}^{f} \mid \tau^{+}\sigma_{-\mu} \mid
p_{2c}^{i}>
\end{split}
\end{equation}

\begin{eqnarray}
<p^{f}n_{c}^f \mid \tau^{+}\sigma_{\mu} \mid n_{1}^{i}n_{2c}^{i}>
 = +\delta (n^{f},n_{2}^{i}) <p_{c}^{f} \mid \tau^{+}\sigma_{\mu} \mid n_{1c}^{i}>\nonumber\\
-\delta (n^{f},n_{1}^{i}) <p_{c}^{f} \mid \tau^{+}\sigma_{\mu} \mid
n_{2c}^{i}>,\nonumber\\
\end{eqnarray}
where $\tau^{+}$ are the isospin raising operators ($\tau^+ \mid p> = \mid n>$) and $\overrightarrow{\sigma}$ is the Pauli  spin operator.
Four-proton (four-neutron) and higher q.p. states were not considered in the model and may be taken as a future task.

For odd-A nuclei low-lying states were obtained by lifting the q.p. in
the orbit of the smallest energy to higher-lying orbits \cite{Mut92}. The excited states were constructed by following three mechanisms\\
(i) lifting the odd neutron (proton) from ground state to excited states,\\
(ii) three-neutron (three-proton) states, corresponding to excitation of a neutron (proton), or,\\
(iii) one-neutron two-proton (one-proton two-neutron) states, corresponding to excitation of
a proton (neutron).

The reduction of multi-q.p. transitions to
correlated ($c$) one-q.p. states were given by,

\begin{flalign}
<p_{1}^{f}n_{1}^{f}n_{2c}^{f} \mid \tau^{+}\sigma_{\mu} \mid n_{1}^{i}n_{2}^{i}n_{3c}^{i}>
 = \delta (n_{1}^{f},n_{2}^{i}) \delta (n_{2}^{f},n_{3}^{i}) \quad\quad \nonumber\\ <p_{1c}^{f} \mid \tau^{+}\sigma_{\mu} \mid n_{1c}^{i}>
 - \delta (n_{1}^{f},n_{1}^{i}) \delta (n_{2}^{f},n_{3}^{i})
 <p_{1c}^{f} \mid \tau^{+}\sigma_{\mu} \quad\nonumber\\ \mid n_{2c}^{i}>
 + \delta (n_{1}^{f},n_{1}^{i}) \delta (n_{2}^{f},n_{2}^{i})<p_{1c}^{f} \mid \tau^{+}\sigma_{\mu} \mid n_{3c}^{i}>\quad\quad
\end{flalign}

\begin{flalign}
<p_{1}^{f}n_{1}^{f}n_{2c}^{f} \mid \tau^{+}\sigma_{-\mu} \mid p_{1}^{i}p_{2}^{i}n_{1c}^{i}>
 = \delta (p_{1}^{f},p_{2}^{i})[ \delta (n_{1}^{f},n_{1}^{i}) \quad\quad\nonumber\\<n_{2c}^{f} \mid \tau^{+}\sigma_{-\mu} \mid p_{1c}^{i}>
 - \delta (n_{2}^{f},n_{1}^{i})
<n_{1c}^{f} \mid \tau^{+}\sigma_{-\mu} \mid p_{1c}^{i}>]\nonumber\\
 -\delta (p_{1}^{f},p_{1}^{i})[ \delta (n_{1}^{f},n_{1}^{i}) <n_{2c}^{f} \mid \tau^{+}\sigma_{-\mu} \mid
 p_{2c}^{i}> \nonumber\\
  - \delta (n_{2}^{f},n_{1}^{i}) <n_{1c}^{f} \mid \tau^{+}\sigma_{-\mu} \mid
  p_{2c}^{i}>]
\end{flalign}

\begin{flalign}
<p_{1}^{f}p_{2}^{f}p_{3c}^{f} \mid \tau^{+}\sigma_{\mu} \mid p_{1}^{i}p_{2}^{i}n_{1c}^{i}>
 = \delta (p_{2}^{f},p_{1}^{i}) \delta (p_{3}^{f},p_{2}^{i})\nonumber\\ <p_{1c}^{f} \mid \tau^{+}\sigma_{\mu} \mid n_{1c}^{i}>
 - \delta (p_{1}^{f},p_{1}^{i}) \delta (p_{3}^{f},p_{2}^{i})
<p_{2c}^{f} \mid \tau^{+}\sigma_{\mu}\nonumber\\ \mid
 n_{1c}^{i}>+ \delta (p_{1}^{f},p_{1}^{i}) \delta (p_{2}^{f},p_{2}^{i})
<p_{3c}^{f} \mid \tau^{+}\sigma_{\mu} \mid n_{1c}^{i}>
\end{flalign}

\begin{eqnarray}
<p_{1}^{f}p_{2}^{f}n_{1c}^{f} \mid \tau^{+}\sigma_{-\mu} \mid p_{1}^{i}p_{2}^{i}p_{3c}^{i}>
 = \delta (p_{1}^{f},p_{2}^{i}) \delta (p_{2}^{f},p_{3}^{i}) <n_{1c}^{f} \nonumber\\\mid \tau^{+}\sigma_{-\mu} \mid p_{1c}^{i}>
 - \delta (p_{1}^{f},p_{1}^{i}) \delta (p_{2}^{f},p_{3}^{i})
<n_{1c}^{f} \mid \tau^{+}\sigma_{-\mu} \mid p_{2c}^{i}>\nonumber\\ + \delta
(p_{1}^{f},p_{1}^{i}) \delta (p_{2}^{f},p_{2}^{i}) <n_{1c}^{f} \mid
\tau^{+}\sigma_{-\mu} \mid p_{3c}^{i}>\nonumber\\
\end{eqnarray}
\begin{eqnarray}
<p_{1}^{f}p_{2}^{f}n_{1c}^{f} \mid \tau^{+}\sigma_{\mu} \mid p_{1}^{i}n_{1}^{i}n_{2c}^{i}>
 = \delta (n_{1}^{f},n_{2}^{i})[ \delta (p_{1}^{f},p_{1}^{i}) <p_{2c}^{f} \nonumber\\\mid \tau^{+}\sigma_{\mu} \mid n_{1c}^{i}>
 - \delta (p_{2}^{f},p_{1}^{i})
<p_{1c}^{f} \mid \tau^{+}\sigma_{\mu} \mid n_{1c}^{i}>] -\delta
(n_{1}^{f},\nonumber\\n_{1}^{i})[ \delta (p_{1}^{f},p_{1}^{i})
<p_{2c}^{f} \mid \tau^{+}\sigma_{\mu} \mid n_{2c}^{i}>
 - \delta (p_{2}^{f},p_{1}^{i}) \nonumber\\<p_{1c}^{f} \mid \tau^{+}\sigma_{\mu} \mid n_{2c}^{i}>]\hspace{1.2in}
\end{eqnarray}
\begin{eqnarray}
<n_{1}^{f}n_{2}^{f}n_{3c}^{f} \mid \tau^{+}\sigma_{-\mu} \mid p_{1}^{i}n_{1}^{i}n_{2c}^{i}>
=\delta (n_{2}^{f},n_{1}^{i}) \delta (n_{3}^{f},n_{2}^{i})
<n_{1c}^{f} \quad\nonumber\\ \mid \tau^{+}\sigma_{-\mu} \mid p_{1c}^{i}> \mbox{}
-\delta (n_{1}^{f},n_{1}^{i}) \delta
(n_{3}^{f},n_{2}^{i})
<n_{2c}^{f} \mid \tau^{+}\sigma_{-\mu} \mid p_{1c}^{i}> \mbox{}
\nonumber\\+\delta (n_{1}^{f},n_{1}^{i}) \delta (n_{2}^{f},n_{2}^{i})
<n_{3c}^{f} \mid \tau^{+}\sigma_{-\mu} \mid p_{1c}^{i}> \quad \quad \quad \quad
\end{eqnarray}

For an odd-odd nucleus the ground state was assumed to be a
proton-neutron q.p. pair state of smallest energy. Low-lying states were expressed in the q.p. picture by
proton-neutron pair states (two-q.p. states) or by states which were
obtained by adding two-proton or two-neutron q.p.'s (four-q.p.
states) \cite{Mut92}. Reduction of two-q.p.
states into correlated ($c$) one-q.p. states was given as
\begin{eqnarray}
<p_{1}^{f}p_{2c}^{f} \mid \tau^{+}\sigma_{\mu} \mid p^{i}n_{c}^{i}>
= \delta(p_{1}^{f},p^{i}) <p_{2c}^{f} \mid \tau^{+}\sigma_{\mu} \mid
n_{c}^{i}> \nonumber\\- \delta(p_{2}^{f},p^{i}) <p_{1c}^{f} \mid
\tau^{+}\sigma_{\mu} \mid n_{c}^{i}> \quad \quad \quad
\end{eqnarray}
\begin{eqnarray}
<n_{1}^{f}n_{2c}^{f} \mid \tau^{+}\sigma_{-\mu} \mid p^{i}n_{c}^{i}>
= \delta(n_{2}^{f},n^{i}) <n_{1c}^{f} \mid \tau^{+}\sigma_{-\mu} \mid
p_{c}^{i}>\nonumber\\ - \delta(n_{1}^{f},n^{i}) <n_{2c}^{f} \mid
\tau^{+}\sigma_{-\mu} \mid p_{c}^{i}> \quad \quad \quad
\end{eqnarray}
while the four-q.p. states were simplified as
\begin{flalign}
<p_{1}^{f}p_{2}^{f}n_{1}^{f}n_{2c}^{f} \mid \tau^{+}\sigma_{-\mu}
\mid p_{1}^{i}p_{2}^{i}p_{3}^{i}n_{1c}^{i}>
=\delta (n_{2}^{f},n_{1}^{i})[ \delta (p_{1}^{f},p_{2}^{i}\hspace{1.5in}\nonumber\\\delta
(p_{2}^{f},p_{3}^{i})
<n_{1c}^{f} \mid \tau^{+}\sigma_{-\mu} \mid p_{1c}^{i}>
-\delta (p_{1}^{f},p_{1}^{i}) \delta (p_{2}^{f},p_{3}^{i})
<n_{1c}^{f}\hspace{1.5in}\nonumber\\ \mid \tau^{+}\sigma_{-\mu} \mid p_{2c}^{i}> +\delta
(p_{1}^{f},p_{1}^{i}) \delta (p_{2}^{f},p_{2}^{i})
<n_{1c}^{f} \mid \tau^{+}\sigma_{-\mu} \mid p_{3c}^{i}>] \hspace{1.5in}\nonumber\\ -\delta
(n_{1}^{f},n_{1}^{i})[ \delta (p_{1}^{f},p_{2}^{i})\delta
(p_{2}^{f},p_{3}^{i})
<n_{2c}^{f} \mid \tau^{+}\sigma_{-\mu} \mid p_{1c}^{i}> \hspace{1.5in}\nonumber\\
-\delta (p_{1}^{f},p_{1}^{i}) \delta (p_{2}^{f},p_{3}^{i})
<n_{2c}^{f} \mid \tau^{+}\sigma_{-\mu} \mid p_{2c}^{i}> +\delta
(p_{1}^{f},p_{1}^{i}) \delta (p_{2}^{f},\hspace{1.5in}\nonumber\\p_{2}^{i})
<n_{2c}^{f}\mid \tau^{+}\sigma_{-\mu} \mid p_{3c}^{i}>]\hspace{3in}
\end{flalign}
\begin{eqnarray}
<p_{1}^{f}p_{2}^{f}p_{3}^{f}p_{4c}^{f} \mid \tau^{+}\sigma_{\mu} \mid
p_{1}^{i}p_{2}^{i}p_{3}^{i}n_{1c}^{i}>
\nonumber\\
=-\delta (p_{2}^{f},p_{1}^{i}) \delta (p_{3}^{f},p_{2}^{i})\delta
(p_{4}^{f},p_{3}^{i})
<p_{1c}^{f} \mid \tau^{+}\sigma_{\mu} \mid n_{1c}^{i}> \nonumber\\
+\delta (p_{1}^{f},p_{1}^{i}) \delta (p_{3}^{f},p_{2}^{i}) \delta
(p_{4}^{f},p_{3}^{i})
<p_{2c}^{f} \mid \tau^{+}\sigma_{\mu} \mid n_{1c}^{i}> \nonumber\\
-\delta (p_{1}^{f},p_{1}^{i}) \delta (p_{2}^{f},p_{2}^{i}) \delta
(p_{4}^{f},p_{3}^{i})
<p_{3c}^{f} \mid \tau^{+}\sigma_{\mu} \mid n_{1c}^{i}> \nonumber\\
+\delta (p_{1}^{f},p_{1}^{i}) \delta (p_{2}^{f},p_{2}^{i}) \delta
(p_{3}^{f},p_{3}^{i}) <p_{4c}^{f} \mid \tau^{+}\sigma_{\mu} \mid
n_{1c}^{i}> \quad
\end{eqnarray}
\begin{eqnarray}
<p_{1}^{f}p_{2}^{f}n_{1}^{f}n_{2c}^{f} \mid \tau^{+}\sigma_{\mu} \mid
p_{1}^{i}n_{1}^{i}n_{2}^{i}n_{3c}^{i}>
=\delta (p_{1}^{f},p_{1}^{i})[ \delta (n_{1}^{f},n_{2}^{i})\nonumber\\\delta
(n_{2}^{f},n_{3}^{i})
<p_{2c}^{f} \mid \tau^{+}\sigma_{\mu} \mid n_{1c}^{i}>
-\delta (n_{1}^{f},n_{1}^{i}) \delta (n_{2}^{f},n_{3}^{i})\nonumber\\
<p_{2c}^{f} \mid \tau^{+}\sigma_{\mu} \mid n_{2c}^{i}> +\delta
(n_{1}^{f},n_{1}^{i}) \delta (n_{2}^{f},n_{2}^{i})
<p_{2c}^{f} \mid \tau^{+}\sigma_{\mu} \nonumber\\\mid n_{3c}^{i}>]  -\delta
(p_{2}^{f},p_{1}^{i})[ \delta (n_{1}^{f},n_{2}^{i})\delta
(n_{2}^{f},n_{3}^{i})\nonumber\\
<p_{1c}^{f} \mid \tau^{+}\sigma_{\mu} \mid n_{1c}^{i}>
-\delta (n_{1}^{f},n_{1}^{i}) \delta (n_{2}^{f},n_{3}^{i})
<p_{1c}^{f} \mid \tau^{+}\sigma_{\mu} \nonumber\\\mid n_{2c}^{i}> +\delta
(n_{1}^{f},n_{1}^{i}) \delta (n_{2}^{f},n_{2}^{i})
<p_{1c}^{f} \mid \tau^{+}\sigma_{\mu} \mid n_{3c}^{i}>]\quad\quad
\end{eqnarray}
\begin{eqnarray}
<n_{1}^{f}n_{2}^{f}n_{3}^{f}n_{4c}^{f} \mid \tau^{+}\sigma_{-\mu}
\mid p_{1}^{i}n_{1}^{i}n_{2}^{i}n_{3c}^{i}>
=\nonumber\\ +\delta (n_{2}^{f},n_{1}^{i}) \delta (n_{3}^{f},n_{2}^{i})\delta
(n_{4}^{f},n_{3}^{i})
<n_{1c}^{f} \mid \tau^{+}\sigma_{-\mu} \mid p_{1c}^{i}> \nonumber\\
-\delta (n_{1}^{f},n_{1}^{i}) \delta (n_{3}^{f},n_{2}^{i}) \delta
(n_{4}^{f},n_{3}^{i})
<n_{2c}^{f} \mid \tau^{+}\sigma_{-\mu} \mid p_{1c}^{i}> \nonumber\\
+\delta (n_{1}^{f},n_{1}^{i}) \delta (n_{2}^{f},n_{2}^{i}) \delta
(n_{4}^{f},n_{3}^{i})
<n_{3c}^{f} \mid \tau^{+}\sigma_{-\mu} \mid p_{1c}^{i}> \nonumber\\
-\delta (n_{1}^{f},n_{1}^{i}) \delta (n_{2}^{f},n_{2}^{i}) \delta
(n_{3}^{f},n_{3}^{i}) <n_{4c}^{f} \mid \tau^{+}\sigma_{-\mu} \mid
p_{1c}^{i}> \quad\quad \label{last}
\end{eqnarray}
For all the given q.p. transition amplitudes [Eqs. ~(\ref{first})- ~(\ref{last})],
the antisymmetrization of the single- q.p. states was taken into account:\\
$ p_{1}^{f}<p_{2}^{f}<p_{3}^{f}<p_{4}^{f}$,\\
$ n_{1}^{f}<n_{2}^{f}<n_{3}^{f}<n_{4}^{f}$,\\
$ p_{1}^{i}<p_{2}^{i}<p_{3}^{i}<p_{4}^{i}$,\\
$ n_{1}^{i}<n_{2}^{i}<n_{3}^{i}<n_{4}^{i}$.\\
GT transitions of phonon excitations for every excited state were
also taken into account. In this case it was assumed that the quasiparticles in
the parent nucleus remained in the same quasiparticle orbits. For  solution of pn-QRPA equation with separable GT forces and construction of reduced transition probabilities please see Ref.~\cite{Mut92}.

\clearpage


 \clearpage



\begin{table}[htbp]
	\centering
	\caption{Comparison of paring gaps calculated by traditional formula (Scheme-I) and 3-point formula (Scheme-II).}
	\begin{tabular}{r|r|r|r}
		
		Nuclei & Scheme-I& \multicolumn{2}{c}{Scheme-II} \\
		\hline
		& $\Delta$n=$\Delta$p & $\Delta$n  & $\Delta$p \\
		$^{45}$Sc & 1.789 & 1.048 & 2.186 \\
		$^{51}$V  & 1.680 & 1.364 & 1.635 \\
		$^{54}$Fe & 1.633 & 1.692 & 1.521 \\
		$^{55}$Fe & 1.618 & 1.495 & 1.254 \\
		$^{56}$Fe & 1.604 & 1.363 & 1.568 \\
		$^{56}$Mn & 1.604 & 1.083 & 1.221 \\
		$^{58}$Mn & 1.576 & 0.897 & 1.152 \\
		$^{58}$Fe & 1.576 & 1.466 & 1.765 \\
		$^{59}$Co & 1.562 & 1.211 & 1.691 \\
		
	\end{tabular}%
	\label{tab:addlabel}%
\end{table}%

\begin {table*}
\caption {Calculated nuclear partition
functions for few $fp$-shell nuclei compared with the earlier
statistical calculations of \cite{Rau00,Rau03} denoted by FRDM.
Numbers in parenthesis indicate stellar temperature in units of
$GK$.} \label{npf1}
\begin{center}
\scalebox{0.9}{\begin{tabular}{ >{\centering\arraybackslash}m{.75in} | >{\centering\arraybackslash}m{.75in}| >{\centering\arraybackslash}m{.75in}| >{\centering\arraybackslash}m{.75in}|>{\centering\arraybackslash}m{.75in}|>{\centering\arraybackslash}m{.75in}|>{\centering\arraybackslash}m{.75in}}

        &            &            &             &           &             &          \\
\bf{Nuclei} & \bf{pnQRPA}& \bf{FRDM}  &  \bf{pnQRPA}& \bf{FRDM} & \bf{pnQRPA} & \bf{FRDM} \\
        &\bf{G(0.01)}&\bf{G(0.01)}&\bf{G(1.5)}  &\bf{G(1.5)}&\bf{G(10.0)}   &\bf{G(10.0)}\\
        &\bf{G(0.5)} &\bf{G(0.5)} &\bf{G(3.0)}    &\bf{G(3.0)}  &\bf{G(30.0)}   &\bf{G(30.0)}\\
\hline
   $^{45}$Sc& 1.03E+00 & 1.12E+00 & 2.08E+00 & 1.50E+00 & 8.49E+00 & 6.58E+00 \\
            &1.93E+00 & 1.38E+00 & 2.41E+00 & 1.80E+00 & 2.11E+04 & 7.51E+03 \\
                    &            &            &             &           &             &          \\
   $^{46}$Sc&  2.26E+00 & 1.00E+00 & 5.25E+01 & 2.45E+00 & 1.84E+02 & 1.26E+01 \\
            &3.10E+01 & 1.45E+00 & 6.52E+01 & 3.76E+00 & 1.02E+05 & 7.51E+03 \\
                    &            &            &             &           &             &          \\
   $^{40}$Ti& 1.00E+00 & 1.00E+00 & 1.00E+00 & 1.00E+00 & 1.91E+00 & 2.42E+00 \\
            &1.00E+00 & 1.00E+00 & 1.00E+00 & 1.00E+00 & 2.10E+03 & 5.75E+03 \\
                    &            &            &             &           &             &          \\
   $^{41}$Ti& 1.00E+00 & 1.00E+00 & 1.62E+00 & 1.02E+00 & 5.17E+00 & 3.98E+00 \\
            &1.24E+00 & 1.00E+00 & 1.79E+00 & 1.15E+00 & 1.57E+04 & 3.74E+03 \\
                    &            &            &             &           &             &          \\
   $^{44}$V&5.00E+00 & 1.00E+00 & 5.71E+00 & 1.19E+00 & 2.48E+01 & 8.15E+00 \\
           &5.13E+00 & 1.03E+00 & 7.01E+00 & 1.55E+00 & 8.84E+04 & 7.78E+03 \\
                   &            &            &             &           &             &          \\
    $^{45}$V&1.00E+00 & 1.00E+00 & 1.90E+01 & 1.83E+00 & 3.31E+01 & 3.57E+00 \\
            &8.45E+00 & 1.34E+00 & 2.37E+01 & 2.16E+00 & 2.22E+04 & 6.32E+03 \\
                    &            &            &             &           &             &          \\
    $^{51}$V &8.00E+00 & 1.00E+00 & 9.89E+00 & 1.06E+00 & 4.16E+01 & 3.45E+00 \\
             &8.42E+00 & 1.00E+00 & 1.32E+01 & 1.24E+00 & 3.11E+04 & 1.72E+04 \\
             &         &          &          &          &          &     \\
    $^{42}$Cr&1.00E+00 & 1.00E+00 & 1.00E+00 & 1.00E+00 & 2.04E+00 & 3.74E+00 \\
             &1.00E+00 & 1.00E+00 & 1.00E+00 & 1.00E+00 & 2.39E+03 & 1.11E+04 \\
        &            &            &             &           &             &          \\
    $^{43}$Cr&1.00E+00 & 1.00E+00 & 1.42E+00 & 1.04E+00 & 5.61E+00 & 4.82E+00 \\
             &1.07E+00 & 1.00E+00 & 1.65E+00 & 1.20E+00 & 1.76E+04 & 6.23E+03 \\
                     &            &            &             &           &             &          \\
    $^{56}$Mn&1.00E+00 & 1.03E+00 & 1.21E+01 & 2.36E+00 & 1.74E+02 & 1.94E+01 \\
              &4.45E+00& 1.43E+00 & 2.80E+01 & 3.98E+00 & 2.20E+05 & 1.73E+05 \\
                      &            &            &             &           &             &          \\
    $^{57}$Mn&1.00E+00 & 1.00E+00 & 2.82E+00 & 1.53E+00 & 3.02E+01 & 1.10E+01 \\
             &1.71E+00 & 1.14E+00 & 3.83E+00 & 1.78E+00 & 4.32E+04 & 2.65E+05 \\
                     &            &            &             &           &             &          \\
    $^{58}$Mn&3.00E+00 & 1.00E+00 & 1.62E+01 & 1.25E+00 & 1.22E+02 & 1.95E+01 \\
             &6.72E+00 & 1.03E+00 & 2.91E+01 & 1.76E+00 & 2.56E+05 & 4.18E+05 \\
             &         &          &          &          &          &         \\
    $^{54}$Fe&1.00E+00 &1.00E+00  & 1.00E+00 & 1.00E+00 & 7.04E+00 & 5.54E+00 \\
             &1.00E+00 & 1.00E+00 & 1.02E+00 & 1.02E+00 & 3.75E+03 & 9.60E+04 \\
             &         &          &          &          &          &     \\
    $^{55}$Fe&4.00E+00 & 1.00E+00 & 4.72E+00 & 1.02E+00 & 3.09E+01 & 6.02E+00 \\
             &4.23E+00 & 1.00E+00 & 5.65E+00 & 1.17E+00 & 3.73E+04 & 6.21E+04 \\
             &         &          &          &          &          &      \\
    $^{58}$Fe&1.00E+00 & 1.00E+00 & 1.01E+00 & 1.01E+00 & 1.25E+01 & 1.09E+01 \\
             &1.00E+00 & 1.00E+00 & 1.23E+00 & 1.23E+00 & 4.47E+03 & 6.99E+05 \\
                 &            &            &             &           &             &          \\
    $^{59}$Co&8.00E+00 & 1.00E+00 & 8.57E+00 & 1.00E+00 & 4.99E+01 & 6.25E+00 \\
             &8.17E+00 & 1.00E+00 & 9.18E+00 & 1.03E+00 & 4.73E+04 & 2.02E+05 \\
                     &            &            &             &           &             &          \\
    $^{56}$Zn&1.00E+00 & 1.00E+00 & 1.00E+00 & 1.00E+00 & 2.84E+00 & 5.29E+00 \\
             &1.00E+00 & 1.00E+00 & 1.00E+00 & 1.00E+00 & 4.04E+03 & 1.85E+05 \\
                     &            &            &             &           &             &          \\
    $^{57}$Zn&8.00E+00 & 1.00E+00 & 8.70E+00 & 1.03E+00 & 1.64E+01 & 4.81E+00 \\
             & 8.31E+00 & 1.00E+00 & 9.02E+00 & 1.16E+00 & 4.01E+04 & 5.96E+04 \\
                     &            &            &             &           &             &          \\
    $^{62}$Ga&1.00E+00 & 1.00E+00 & 1.40E+00 & 3.98E+00 & 4.73E+01 & 2.33E+02 \\
             &1.02E+00 & 1.41E+00 & 3.16E+00 & 1.01E+01 & 3.56E+05 & 7.26E+06 \\
                     &            &            &             &           &             &          \\
    $^{63}$Ga&4.00E+00 & 1.00E+00 & 7.66E+00 & 1.26E+00 & 3.41E+01 & 2.54E+01 \\
             &5.04E+00 & 1.04E+00 & 1.07E+01 & 1.79E+00 & 6.11E+04 & 1.96E+06 \\
                     &            &            &             &           &             &          \\
    $^{59}$Ge&1.00E+00 & 1.00E+00 & 1.77E+00 & 1.05E+00 & 1.25E+01 & 7.01E+00 \\
             &1.44E+00 & 1.00E+00 & 2.03E+00 & 1.20E+00 & 4.94E+04 & 1.30E+05 \\
                     &            &            &             &           &             &          \\
    $^{60}$Ge&1.00E+00 & 1.00E+00 & 1.00E+00 & 1.00E+00 & 3.61E+00 & 6.04E+00 \\
             &1.00E+00 & 1.00E+00 & 1.00E+00 & 1.00E+00 & 4.53E+03 & 5.67E+05 \\
                     &            &            &             &           &             &          \\
\hline
\end {tabular}}
\end{center}
\end {table*}

\begin {table*}[h]
\caption {Same as Tab.~\ref{npf1}.} \label{npf2}
\begin{center}
\scalebox{0.9}{\begin{tabular}{ >{\centering\arraybackslash}m{.75in} | >{\centering\arraybackslash}m{.75in}| >{\centering\arraybackslash}m{.75in} |>{\centering\arraybackslash}m{.75in}|>{\centering\arraybackslash}m{.75in}|>{\centering\arraybackslash}m{.75in}|>{\centering\arraybackslash}m{.75in}}
        &            &            &             &           &             &           \\
\bf{Nuclei} & \bf{pnQRPA}& \bf{FRDM}  &  \bf{pnQRPA}& \bf{FRDM} & \bf{pnQRPA} & \bf{FRDM} \\
        &\bf{G(0.01)}&\bf{G(0.01)}&\bf{G(1.5)}  &\bf{G(1.5)}&\bf{G(10.0)}   &\bf{G(10.0)}\\
        &\bf{G(0.5)} &\bf{G(0.5)} &\bf{G(3.0)}    &\bf{G(3.0)}  &\bf{G(30.0)}   &\bf{G(30.0)}\\
\hline

    $^{65}$As&1.00E+00 & 1.00E+00 & 1.26E+00 & 1.31E+00 & 1.56E+01 & 3.13E+01 \\
             &1.02E+00 & 1.04E+00 & 1.64E+00 & 1.92E+00 & 6.80E+04 & 2.82E+06 \\
                     &            &            &             &           &             &          \\
    $^{66}$As&1.00E+00 & 1.00E+00 & 2.17E+00 & 1.83E+00 & 5.75E+01 & 9.45E+01 \\
             &1.36E+00 & 1.11E+00 & 3.47E+00 & 3.61E+00 & 4.51E+05 & 7.49E+06 \\
                     &            &            &             &           &             &          \\
    $^{65}$Se&4.00E+00 & 1.00E+00 & 4.60E+00 & 1.80E+00 & 2.19E+01 & 7.07E+01 \\
             &4.14E+00 & 1.11E+00 & 5.12E+00 & 3.38E+00 & 6.67E+04 & 7.39E+06 \\
                     &            &            &             &           &             &          \\
    $^{66}$Se&1.00E+00 & 1.00E+00 & 1.00E+00 & 1.00E+00 & 4.91E+00 & 2.71E+01 \\
             &1.00E+00 & 1.00E+00 & 1.00E+00 & 1.17E+00 & 5.82E+03 & 1.11E+07 \\
                   &            &            &             &           &             &           \\
  $^{69}$Br&1.00E+00 & 1.00E+00 & 1.33E+00 & 1.19E+00 & 1.61E+01 & 2.03E+01 \\
             &1.03E+00 & 1.03E+00 & 1.70E+00 & 1.57E+00 & 7.33E+04 & 4.07E+06 \\
        &            &            &             &           &             &           \\
    $^{70}$Br&1.00E+00 & 1.00E+00 & 4.44E+00 & 1.60E+00 & 7.62E+01 & 7.37E+01 \\
             &2.43E+00 & 1.08E+00 & 6.68E+00 & 2.87E+00 & 4.13E+05 & 1.10E+07 \\
                     &            &            &             &           &             &           \\
    $^{69}$Kr&1.00E+00 & 1.00E+00 & 1.55E+00 & 2.53E+00 & 1.92E+01 & 1.32E+02 \\
             &1.14E+00 & 1.21E+00 & 1.95E+00 & 5.53E+00 & 7.86E+04 & 3.00E+07 \\
                     &            &            &             &           &             &           \\
    $^{70}$Kr&1.00E+00 & 1.00E+00 & 1.00E+00 & 1.00E+00 & 5.39E+00 & 4.49E+01 \\
             &1.00E+00 & 1.00E+00 & 1.00E+00 & 2.01E+00 & 6.31E+03 & 3.80E+07 \\
                     &            &            &             &           &             &           \\
    $^{74}$Rb&1.00E+00 & 1.00E+00 & 2.91E+00 & 9.14E+00 & 9.48E+01 & 1.17E+03 \\
             &1.33E+00 & 2.10E+00 & 6.19E+00 & 2.64E+01 & 3.74E+05 & 4.02E+08 \\
                     &            &            &             &           &             &           \\
    $^{75}$Rb&4.00E+00 & 1.00E+00 & 1.01E+01 & 1.64E+00 & 8.02E+01 & 8.79E+01 \\
             &5.87E+00 & 1.09E+00 & 1.81E+01 & 2.94E+00 & 1.00E+05 & 8.28E+07 \\
                     &            &            &             &           &             &           \\
    $^{73}$Sr&4.00E+00 & 1.00E+00 & 5.30E+00 & 2.13E+00 & 2.62E+01 & 1.16E+02 \\
             &4.39E+00 & 1.16E+00 & 6.02E+00 & 4.35E+00 & 9.40E+04 & 5.42E+07 \\
                     &            &            &             &           &             &           \\
    $^{74}$Sr&1.00E+00 & 1.00E+00 & 1.00E+00 & 1.26E+00 & 5.64E+00 & 6.59E+01 \\
             &1.00E+00 & 1.00E+00 & 1.00E+00 & 2.46E+00 & 7.14E+03 & 1.24E+08 \\
                     &            &            &             &           &             &           \\
    $^{78}$Y &1.00E+00 & 1.00E+00 & 2.38E+00 & 1.97E+00 & 1.06E+02 & 1.67E+02 \\
             &1.08E+00 & 1.13E+00 & 6.90E+00 & 4.05E+00 & 3.30E+05 & 1.01E+08 \\
                     &            &            &             &           &             &           \\
    $^{79}$Y &6.00E+00 & 1.00E+00 & 8.41E+00 & 1.35E+00 & 6.12E+01 & 6.35E+01 \\
             &6.11E+00 & 1.05E+00 & 1.32E+01 & 2.07E+00 & 1.24E+05 & 1.34E+08 \\
                     &            &            &             &           &             &           \\
    $^{80}$Zr&1.00E+00 & 1.00E+00 & 1.55E+00 & 1.55E+00 & 1.67E+01 & 7.46E+01 \\
             &1.01E+00 & 1.01E+00 & 3.03E+00 & 3.30E+00 & 8.58E+03 & 5.23E+08 \\
                     &            &            &             &           &             &           \\
    $^{81}$Zr&4.00E+00 & 1.00E+00 & 8.19E+00 & 1.01E+00 & 8.10E+01 & 1.55E+02 \\
             &4.26E+00 & 1.00E+00 & 1.61E+01 & 1.24E+00 & 1.34E+05 & 7.31E+08 \\
                     &            &            &             &           &             &           \\
    $^{82}$Nb&1.00E+00 & 1.00E+00 & 3.36E+00 & 2.89E+00 & 1.09E+02 & 5.56E+02 \\
             &2.13E+00 & 1.25E+00 & 5.36E+00 & 7.25E+00 & 3.08E+05 & 6.99E+08 \\
                     &            &            &             &           &             &           \\
    $^{83}$Nb&6.00E+00 & 1.00E+00 & 1.22E+01 & 1.91E+00 & 8.68E+01 & 6.91E+01 \\
             &6.82E+00 & 1.07E+00 & 2.04E+01 & 3.26E+00 & 1.33E+05 & 4.74E+08 \\
                     &            &            &             &           &             &           \\
    $^{83}$Mo&1.00E+00 & 1.00E+00 & 2.21E+00 & 1.21E+00 & 4.03E+01 & 1.25E+02 \\
             &1.38E+00 & 1.03E+00 & 2.94E+00 & 1.72E+00 & 1.26E+05 & 4.40E+08 \\
                     &            &            &             &           &             &           \\
    $^{84}$Mo&1.00E+00 & 1.00E+00 & 1.16E+00 & 1.00E+00 & 1.65E+01 & 2.78E+02 \\
             &1.00E+00 & 1.00E+00 & 2.02E+00 & 1.02E+00 & 8.68E+03 & 3.64E+09 \\
                     &            &            &             &           &             &           \\
    $^{86}$Tc&1.00E+00 & 1.00E+00 & 1.49E+00 & 3.69E+00 & 1.52E+01 & 1.04E+03 \\
             &1.08E+00 & 1.33E+00 & 2.58E+00 & 1.08E+01 & 1.36E+04 & 1.95E+09 \\
                     &            &            &             &           &             &           \\
    $^{87}$Tc&1.00E+01 & 1.00E+00 & 1.05E+01 & 1.08E+00 & 2.13E+01 & 2.45E+02 \\
             &1.01E+01 & 1.00E+00 & 1.16E+01 & 1.65E+00 & 4.73E+03 & 1.68E+09 \\
\hline
\end {tabular}}
\end{center}
\end {table*}

\begin{table}[htbp]
	\centering
	\caption{Comparison of calculated ECC for different pairing gaps at stellar temperature of 5.8 $GK$.}
	\begin{tabular}{rrrrrrrrrr}
		
		\multicolumn{1}{c}{Nuclei} & Pairing gap & w(MeV) = 2 & w(MeV) = 3 & w(MeV) = 5 & w(MeV) = 10 & w(MeV) = 15 & w(MeV) = 20 & w(MeV) = 25 & w(MeV) = 30 \\
		\hline
		$^{45}$Sc &Scheme-I & 1.05E-49 & 1.50E-48 & 1.70E-47 & 3.77E-45 & 1.93E-43 & 2.61E-42 & 1.84E-41 & 8.80E-41 \\
		& Scheme-II & 6.33E-50 & 8.39E-49 & 7.68E-48 & 3.18E-45 & 1.60E-43 & 2.12E-42 & 1.48E-41 & 7.02E-41 \\
		$^{51}$V &Scheme-I & 5.27E-50 & 1.04E-48 & 3.05E-47 & 5.21E-45 & 1.43E-43 & 1.54E-42 & 9.75E-42 & 4.37E-41 \\
		& Scheme-II & 4.37E-50 & 8.74E-49 & 2.75E-47 & 5.13E-45 & 1.41E-43 & 1.52E-42 & 9.54E-42 & 4.26E-41 \\
		$^{54}$Fe &Scheme-I & 3.52E-49 & 6.74E-48 & 1.92E-46 & 3.55E-44 & 1.03E-42 & 1.14E-41 & 7.27E-41 & 3.28E-40 \\
		& Scheme-II & 3.71E-49 & 7.12E-48 & 2.00E-46 & 3.52E-44 & 1.01E-42 & 1.12E-41 & 7.15E-41 & 3.23E-40 \\
		$^{55}$Fe  &Scheme-I & 4.23E-49 & 7.34E-48 & 1.29E-46 & 6.05E-45 & 2.22E-43 & 3.19E-42 & 2.37E-41 & 1.18E-40 \\
		& Scheme-II & 3.58E-49 & 6.10E-48 & 1.03E-46 & 5.60E-45 & 2.28E-43 & 3.25E-42 & 2.40E-41 & 1.18E-40 \\
		$^{56}$Fe  &Scheme-I & 1.68E-49 & 4.51E-48 & 2.12E-46 & 4.42E-44 & 1.09E-42 & 1.07E-41 & 6.37E-41 & 2.73E-40 \\
		& Scheme-II & 1.56E-49 & 4.35E-48 & 2.14E-46 & 4.62E-44 & 1.14E-42 & 1.12E-41 & 6.67E-41 & 2.86E-40 \\
		$^{56}$Mn  &Scheme-I & 2.82E-50 & 4.27E-49 & 6.39E-48 & 1.42E-45 & 6.10E-44 & 7.87E-43 & 5.43E-42 & 2.56E-41 \\
		& Scheme-II & 2.06E-50 & 2.96E-49 & 5.26E-48 & 2.11E-45 & 8.34E-44 & 1.02E-42 & 6.86E-42 & 3.18E-41 \\
		$^{58}$Mn  &Scheme-I & 7.38E-51 & 1.83E-49 & 1.38E-47 & 5.61E-45 & 1.70E-43 & 1.85E-42 & 1.16E-41 & 5.13E-41 \\
		& Scheme-II & 2.92E-51 & 9.64E-50 & 1.06E-47 & 4.70E-45 & 1.42E-43 & 1.54E-42 & 9.61E-42 & 4.25E-41 \\
		$^{58}$Fe  &Scheme-I & 3.11E-49 & 8.52E-48 & 3.71E-46 & 6.03E-44 & 1.27E-42 & 1.15E-41 & 6.42E-41 & 2.64E-40 \\
		& Scheme-II & 3.01E-49 & 8.27E-48 & 3.60E-46 & 5.86E-44 & 1.24E-42 & 1.11E-41 & 6.22E-41 & 2.56E-40 \\
		$^{59}$Co  &Scheme-I & 6.92E-50 & 1.22E-48 & 2.54E-47 & 3.12E-45 & 1.02E-43 & 1.23E-42 & 8.34E-42 & 3.90E-41 \\
		& Scheme-II & 4.66E-50 & 7.91E-49 & 1.62E-47 & 2.75E-45 & 9.66E-44 & 1.17E-42 & 7.89E-42 & 3.67E-41 \\
		
	\end{tabular}%
	\label{tab:addlabel}%
\end{table}%

\begin{table}[htbp]
  \centering
  \caption{Comparison of EC rates for selected nuclei at three temperatures (5, 25 and 30 GK) at three densities (10$^1$g/cm$^3$, 10$^4$g/cm$^3$ and 10$^8$g/cm$^3$) with respect to different pairing gaps.}
    \begin{tabular}{r|r|rr|rr|rr}

    \multicolumn{1}{c}{Nuclei} & \multicolumn{1}{c}{T (GK)} & \multicolumn{2}{c}{10$^1$g/cm$^3$} & \multicolumn{2}{c}{10$^4$g/cm$^3$} & \multicolumn{2}{c}{10$^8$g/cm$^3$}  \\
    \hline
          &       &Scheme-I& Scheme-II & Scheme-I& Scheme-II & Scheme-I &Scheme-II    \\
    $^{45}$Sc & 5     & -4.243 & -4.132 & -4.241 & -4.131 & -2.375 & -2.257   \\
          & 25    & 1.643 & 1.653 & 1.643 & 1.653 & 1.703 & 1.713   \\
          & 30    & 2.268 & 2.256 & 2.268 & 2.257 & 2.303 & 2.292   \\
    \hline
    $^{51}$V & 5     & -5.111 & -5.103 & -5.110 & -5.102 & -2.974 & -2.963   \\
          & 25    & 1.873 & 1.886 & 1.873 & 1.887 & 1.934 & 1.947  \\
          & 30    & 2.531 & 2.538 & 2.531 & 2.539 & 2.566 & 2.574  \\
    \hline
    $^{54}$Fe & 5     & -3.400  & -3.432 & -3.399 & -3.431 & -1.275 & -1.308  \\
          & 25    & 2.059 & 2.054 & 2.059 & 2.055 & 2.118 & 2.114   \\
          & 30    & 2.516 & 2.511 & 2.516 & 2.512 & 2.55  & 2.546   \\
    \hline
    $^{55}$Fe  & 5     & -2.627 & -2.638 & -2.626 & -2.637 & -0.730 & -0.743   \\
          & 25    & 2.305 & 2.358 & 2.306 & 2.359 & 2.366 & 2.419   \\
          & 30    & 2.935 & 2.983 & 2.935 & 2.983 & 2.970  & 3.018   \\
    \hline
    $^{56}$Fe  & 5     & -5.784 & -5.768 & -5.782 & -5.767 & -3.703 & -3.694 \\
          & 25    & 1.718 & 1.742 & 1.719 & 1.742 & 1.779 & 1.802  \\
          & 30    & 2.249 & 2.277 & 2.250  & 2.277 & 2.284 & 2.312  \\
    \hline
    $^{56}$Mn  & 5     & -6.102 & -5.585 & -6.101 & -5.584 & -3.901 & -3.400   \\
          & 25    & 2.080  & 2.392 & 2.081 & 2.393 & 2.141 & 2.453  \\
          & 30    & 2.766 & 3.037 & 2.766 & 3.038 & 2.801 & 3.073  \\
    \hline
    $^{58}$Mn & 5     & -7.217 & -7.197 & -7.216 & -7.195 & -5.007 & -4.988  \\
          & 25    & 2.137 & 2.230  & 2.138 & 2.230  & 2.198 & 2.290   \\
          & 30    & 2.823 & 2.918 & 2.823 & 2.919 & 2.858 & 2.954  \\
    \hline
    $^{58}$Fe  & 5     & -7.424 & -7.426 & -7.423 & -7.424 & -5.242 & -5.245  \\
          & 25    & 1.432 & 1.455 & 1.432 & 1.455 & 1.492 & 1.515  \\
          & 30    & 2.009 & 2.031 & 2.009 & 2.031 & 2.044 & 2.066  \\
    \hline
    $^{59}$Co   & 5     & -5.187 & -5.260 & -5.186 & -5.258 & -3.066 & -3.146  \\
          & 25    & 1.755 & 1.774 & 1.755 & 1.775 & 1.816 & 1.835  \\
          & 30    & 2.453 & 2.471 & 2.453 & 2.472 & 2.488 & 2.507  \\

    \end{tabular}%
  \label{tab:addlabel}%
\end{table}%

\end{document}